\documentclass[12pt]{article}
\usepackage{amssymb,amsmath,epsfig}
\allowdisplaybreaks

\begin{document}

\title{\bf Study of Decoupled Anisotropic Solutions in $f(R,T,R_{\rho\eta}T^{\rho\eta})$ Theory}
\author{Tayyab Naseer \thanks{tayyabnaseer48@yahoo.com; tayyab.naseer@math.uol.edu.pk}
~and M. Sharif \thanks{msharif.math@pu.edu.pk} \\
Department of Mathematics, University of the Punjab,\\
Quaid-i-Azam Campus, Lahore-54590, Pakistan.}

\date{}
\maketitle

\begin{abstract}
In this paper, we consider isotropic solution and extend it to two
different exact well-behaved spherical anisotropic solutions through
minimal geometric deformation method in
$f(R,T,R_{\rho\eta}T^{\rho\eta})$ gravity. We only deform the radial
metric component that separates the field equations into two sets
corresponding to their original sources. The first set corresponds
to perfect matter distribution while the other set exhibits the
effects of additional source, i.e., anisotropy. The isotropic system
is resolved by assuming the metric potentials proposed by
Krori-Barua while the second set needs one constraint to be solved.
The physical acceptability and consistency of the obtained solutions
are analyzed through graphical analysis of effective matter
components and energy bounds. We also examine mass, surface redshift
and compactness of the resulting solutions. For particular values of
the decoupling parameter, our both solutions turn out to be viable
and stable. We conclude that this curvature-matter coupling gravity
provides more stable solutions corresponding to a self-gravitating
geometry.
\end{abstract}
{\bf Keywords:} $f(R,T,R_{\rho\eta}T^{\rho\eta})$ gravity;
Anisotropy; Gravitational decoupling; Self-gravitating systems. \\
{\bf PACS:} 04.50.Kd ; 04.40.Dg;  04.40.-b.

\section{Introduction}

Einstein theory of general relativity (GR) has been considered as
the root of cosmology and gravitational phenomena. Cosmological
findings show that the astronomical objects are not scattered
randomly in the universe but are organized in a systematic way. The
investigation of this arrangement and physical characteristics of
interstellar bodies enable us to figure out accelerated expansion of
the cosmos. This expansion is presumed to be performed by an obscure
form of energy known as dark energy. Moreover, the virial mass
discrepancy at the galactic cluster level and the galaxy rotation
curves \cite{1,1a}, cosmic accelerated expansion as well as other
cosmological observations suggest that the standard general
relativistic gravitational field equations, based on the
Einstein-Hilbert (EH) action cannot describe the universe at large
scales. From cosmological point of view, dark matter and dark energy
components are introduced by hand, in addition to ordinary matter
and energy in this theory. The modifications to GR are found to be
crucial in unveiling mysterious aspects of our universe. The $f(R)$
theory is the immediate extension of GR, formulated on the basis of
an arbitrary function that replaces the Ricci scalar $R$ in the EH
action. The stability of $f(R)$ theory has been discussed by various
researchers by using different approaches \cite{2}-\cite{2b}.
Capozziello et al. \cite{8} studied the stability of different stars
in $f(R)$ theory by utilizing the Lan\'{e}-Emden equation. Recently,
various experiments have been conducted on the astronomical objects
to discuss their composition and stability in this theory
\cite{9}-\cite{9g}.

Later, Bertolami et al. \cite{10} considered the Lagrangian
depending on scalar curvature $R$ and $\mathcal{L}_{m}$ to study the
effects of coupling in $f(R)$ gravity. The coupling between matter
and spacetime in extended theories of GR has encouraged several
theorists to focus on cosmic accelerated expansion. Harko et al.
\cite{20} proposed $f(R,T)$ theory to study the non-minimal
interaction between matter and geometry, $T$ represents trace of the
energy-momentum tensor (EMT). It has been observed that such a
coupling results in the non-conservation of EMT which may cause the
accelerated interstellar expansion. Haghani et al. \cite{22}
presented a wider and more complex theory by adding an extra term in
the Lagrangian of $f(R,T)$ theory to study the strong effects of
non-minimal coupling, referred to $f(R,T,Q)$ theory, in which
$Q\equiv R_{\rho\eta}T^{\rho\eta}$. Indeed, examples of such
couplings can be found in the Einstein-Born-Infeld theories when one
expands the square root in the Lagrangian. In this framework, Sharif
and Zubair investigated the energy bounds for some particular models
\cite{22a} and checked the feasibility of thermodynamical laws
\cite{22b}.

This theory was constructed on the basis of insertion of the strong
non-minimal matter-geometry coupling which is explained by the
factor $Q$. The role of dark matter and dark energy, without
resorting to exotic matter distribution is explained through the
modification in the EH action. Several extended theories such as
$f(R,\mathcal{L}_m)$ and $f(R,T)$ also engage such arbitrary
coupling but their functionals cannot be considered in the most
general form to understand the effects of coupling on celestial
objects in some situations. It should be pointed out that the factor
$R_{\rho\eta}T^{\rho\eta}$ could interpret non-minimal interaction
in the scenario where $f(R,T)$ theory fails to describe. In
particular, one cannot explain coupling effects on the gravitational
model in $f(R,T)$ theory when trace-free EMT (i.e., $T=0$) is
considered, while $f(R,T,Q)$ gravity studies such effects even in
this context. This theory was shown to be stable against
Dolgov-Kawasaki instability and can help to explain the galactic
rotation curves due to the presence of an additional force which
stops the motion of test particles in geodesic path. Haghani et al.
\cite{22} discussed cosmological applications of three different
models in this framework, i.e., $R+\alpha Q,~R(1+\alpha Q)$ and
$R+\beta\sqrt{\mid T\mid}+\alpha Q$, where $\alpha$ and $\beta$ are
arbitrary coupling constants. They analyzed the evolution and
dynamics of the universe for the above models with and without
energy conservation.

Odintsov and S\'{a}ez-G\'{o}mez \cite{23} found some analytical as
well as numerical solutions in $f(R,T,Q)$ theory and compared them
with the $\Lambda$CDM model. They also discussed some problems
related to the instability of fluid distribution. Ayuso et al.
\cite{24} inspected the consistency and reliability of this
complicated theory by choosing some suitable scalar (or vector)
fields. Baffou et al. \cite{25} explored the power-law solution to
understand the early cosmic evolution and checked the stability for
some specific models. Sharif and Waseem \cite{25a,25b} studied
certain physical attributes of massive isotropic/anisotropic
configured stars and checked their stable regions. Yousaf et al.
\cite{26}-\cite{26e} computed several structure scalars for static
and non-static cases which are related with the fundamental
properties of matter distribution. These scalars help to illustrate
the composition and expansion of self-gravitating stellar
configuration.

Owing to the inclusion of highly non-linear terms in the field
equations of a compact geometry, the development of exact solutions
has always been a serious but interesting issue. Gravitational
decoupling is a recently proposed scheme which is used to find
feasible solutions corresponding to the matter distribution
involving multiple sources, such as anisotropy, heat dissipation and
shear stress. The minimal geometric deformation (MGD) technique has
shown significant consequences to achieve physically well-behaved
solutions. This approach offers a variety of enticing ingredients
for new exact solutions for both cosmology and astrophysics. Ovalle
\cite{29} initially proposed this technique to acquire analytical
solutions of stellar objects in the context of braneworld. Later,
Ovalle and Linares \cite{30} found exact spherical isotropic
solutions and concluded that these results are compatible with the
Tolman-IV solution in the braneworld. Casadio et al. \cite{31}
formed the outer spherical solutions and noticed that these
solutions contain singularity at Schwarzschild radius.

Ovalle \cite{32} determined anisotropic solutions via gravitational
decoupling approach. Ovalle et al. \cite{33} extended the isotropic
solutions through this approach and checked the graphical behavior
of new solutions which contain effects of anisotropy. Sharif and
Sadiq \cite{34} developed anisotropic solutions for charged
spherical geometry by taking the Krori-Barua solution and analyzed
the influence of charge on their viability as well as stability.
Sharif and his collaborators \cite{35}-\cite{35c} generalized this
work to $f(G)$ and $f(R)$ theories. Gabbanelli et al. \cite{36}
determined different anisotropic solutions in view of the
Duragpal-Fuloria isotropic spacetime and found them physically
acceptable. Estrada and Tello-Ortiz \cite{36a} constructed various
anisotropic physically consistent solutions by applying this
technique to Heintzmann solution. By taking an appropriate
deformation function, Singh et al. \cite{37} employed embedding
technique to develop anisotropic solutions via this approach. Hensh
and Stuchlik \cite{37a} deformed Tolman VII solution and found
physically feasible anisotropic solutions. Sharif and
Ama-Tul-Mughani \cite{37b,37d} used this technique to find
anisotropic solutions by considering the charged isotropic solution.
Sharif and Majid \cite{37c}-\cite{37f} considered different known
isotropic solutions and found anisotropic spherical solutions with
the help of minimal and extended version of the decoupling scheme in
Brans-Dicke theory.

This paper investigates the influence of $f(R,T,Q)$ correction terms
on two anisotropic solutions obtained through MGD approach for
spherical spacetime. The paper is structured as follows. The basic
formulation of this gravity is presented in the following section.
Section 3 discusses the MGD technique which helps to separate the
gravitational field equations into two sets which correspond to
isotropic and anisotropic configurations. In section 4, we consider
the Krori-Barua spacetime to find new analytic solutions. We also
discuss physical feasibility of the developed anisotropic solutions.
Finally, we summarize our results in the last section.

\section{The $f(R,T,Q)$ Theory}

The corresponding Einstein-Hilbert action is \cite{23}
\begin{equation}\label{g1}
S=\int \frac{1}{16\pi}\left[f(R,T,R_{\rho\eta}T^{\rho\eta})
+\mathcal{L}_{m}\right]\sqrt{-g}d^{4}x,
\end{equation}
where $\mathcal{L}_{m}$ denotes the matter Lagrangian which in this
case is considered to be negative of the energy density of fluid and
$g$ describes determinant of the metric tensor. By adding the
Lagrangian $\mathcal{L}_{\Theta}$, which corresponds to an
additional source term coupled with gravity in the action \eqref{g1}
and varying it with respect to the metric tensor, the field
equations can be written as
\begin{equation}\label{g2}
G_{\rho\eta}=8\pi T_{\rho\eta}^{(tot)},
\end{equation}
where $G_{\rho\eta}$ is the Einstein tensor and the EMT for matter
distribution is
\begin{equation}\label{g3}
T_{\rho\eta}^{(tot)}=T_{\rho\eta}^{(eff)}+\sigma
\Theta_{\rho\eta}=\frac{1}{f_{R}-\mathcal{L}_{m}f_{Q}}T_{\rho\eta}+T_{\rho\eta}^{(D)}+\sigma
\Theta_{\rho\eta},
\end{equation}
$\sigma$ represents the decoupling parameter, $\Theta_{\rho\eta}$
may contain some new fields that produce anisotropic effects in
self-gravitating structure. Also, we can stress
$T_{\rho\eta}^{(eff)}$ as the EMT in $f(R,T,Q)$ gravity which
contains usual as well as modified correction terms. In this case,
the value of $T_{\rho\eta}^{(D)}$ becomes
\begin{eqnarray}
\nonumber
T_{\rho\eta}^{(D)}&=&\frac{1}{8\pi(f_{R}-\mathcal{L}_{m}f_{Q})}
\left[\left(f_{T}+\frac{1}{2}Rf_{Q}\right)T_{\rho\eta}
+\left\{\frac{R}{2}(\frac{f}{R}-f_{R})-\mathcal{L}_{m}f_{T}\right.\right.\\\nonumber
&-&\left.\frac{1}{2}\nabla_{\pi}\nabla_{\beta}(f_{Q}T^{\pi\beta})\right\}g_{\rho\eta}
-\frac{1}{2}\Box(f_{Q}T_{\rho\eta})-(g_{\rho\eta}\Box-
\nabla_{\rho}\nabla_{\eta})f_{R}\\\label{g4}
&-&2f_{Q}R_{\pi(\rho}T_{\eta)}^{\pi}+\nabla_{\pi}\nabla_{(\rho}[T_{\eta)}^{\pi}f_{Q}]
+2(f_{Q}R^{\pi\beta} +\left.f_{T}g^{\pi\beta})\frac{\partial^2
\mathcal{L}_{m}}{\partial g^{\rho\eta}\partial g^{\pi\beta}}\right],
\end{eqnarray}
where $f_{R}=\frac{\partial f(R,T,Q)}{\partial
R},~f_{T}=\frac{\partial f(R,T,Q)}{\partial T},~f_{Q}=\frac{\partial
f(R,T,Q)}{\partial Q}$ and $\nabla_\nu$ describes the covariant
derivative. Also, $\Box\equiv g^{\rho\eta}\nabla_\rho\nabla_\eta$.
The EMT for perfect fluid has the following form
\begin{equation}\label{g5}
T_{\rho\eta}=(\mu+P) u_{\rho} u_{\eta}+Pg_{\rho\eta},
\end{equation}
where $u_{\rho}$ and $P$ are the four-velocity and isotropic
pressure, respectively. In GR, the trace of EMT provides a
particular relationship between $R$ and $T$. One can establish the
trace of $f(R,T,Q)$ field equations as
\begin{align}\nonumber
&3\nabla^{\eta}\nabla_{\eta}
f_R+R\left(f_R-\frac{T}{2}f_Q\right)-T(f_T+1)+\frac{1}{2}
\nabla^{\eta}\nabla_{\eta}(f_QT)+\nabla_\eta\nabla_\rho
(f_QT^{\rho\eta})\\\nonumber
&-2f+(Rf_Q+4f_T)\mathcal{L}_m+2R_{\rho\eta}T^{\rho\eta}f_Q
-2g^{\pi\beta} \frac{\partial^2\mathcal{L}_m}{\partial
g^{\pi\beta}\partial
g^{\rho\eta}}\left(f_Tg^{\rho\eta}+f_QR^{\rho\eta}\right)=0.
\end{align}
The $f(R,T)$ gravity can be achieved from above equation by taking
$Q=0$, while one can also attain $f(R)$ theory for the vacuum case.

The geometry under consideration is distinguished by a hypersurface
$\Sigma$ which delineates the inner and outer sectors of spherical
spacetime. We define the spherical geometry which represents the
interior spacetime as
\begin{equation}\label{g6}
ds^2=-e^{\nu} dt^2+e^{\chi} dr^2+r^2d\theta^2+r^2\sin^2\theta
d\vartheta^2,
\end{equation}
where $\nu=\nu(r)$ and $\chi=\chi(r)$. The corresponding
four-velocity and four-vector in the radial direction are
\begin{equation}\label{g7}
u^\rho=(e^{\frac{-\nu}{2}},0,0,0),\quad w^\rho=(0,e^{\frac{-\chi}{2}},0,0),
\end{equation}
which satisfy the relations $w^\rho u_{\rho}=0, u^\rho u_{\rho}=-1$.
The field equations are
\begin{align}\label{g8}
&e^{-\chi}\left(\frac{\chi'}{r}-\frac{1}{r^2}\right)
+\frac{1}{r^2}=8\pi\left(\mu^{(eff)}-T_{0}^{0(D)}-\sigma
\Theta_{0}^{0}\right),\\\label{g9}
&e^{-\chi}\left(\frac{1}{r^2}+\frac{\nu'}{r}\right)
-\frac{1}{r^2}=8\pi\left(P^{(eff)}+T_{1}^{1(D)}+\sigma
\Theta_{1}^{1}\right),
\\\label{g10}
&-\frac{e^{-\chi}}{4}\left[\chi'\nu'-\nu'^2-2\nu''+\frac{2\chi'}{r}-\frac{2\nu'}{r}\right]
=8\pi\left(P^{(eff)}+T_{2}^{2(D)}+\sigma \Theta_{2}^{2}\right),
\end{align}
where $\mu^{(eff)}=\frac{1}{(f_{R}+\mu f_{Q})}\mu$ and
$P^{(eff)}=\frac{1}{(f_{R}+\mu f_{Q})}P$. Also,
$T_{0}^{0(D)},~T_{1}^{1(D)}$ and $T_{2}^{2(D)}$ represent the
$f(R,T,Q)$ correction terms and make the field equations more
complex. These components are given in Appendix \textbf{A}. Here,
prime means $\frac{\partial}{\partial r}$.

The EMT in this theory, unlike GR and $f(R)$, has non-zero
divergence due to curvature-matter coupling that contributes to
violation of the equivalence principle. Therefore, in the
gravitational field, moving particles do not follow geodesic path
due to the extra force that acts on these particles. Thus we obtain
\begin{align}\nonumber
\nabla^\rho
T_{\rho\eta}&=\frac{2}{2f_T+Rf_Q+16\pi}\left[\nabla_\eta(\mathcal{L}_mf_T)
+\nabla_\rho(f_QR^{\pi\rho}T_{\pi\eta})-G_{\rho\eta}\nabla^\rho(f_Q\mathcal{L}_m)\right.\\\label{g11}
&-\left.\frac{1}{2} (f_Tg_{\pi\beta}+f_QR_{\pi\beta})\nabla_\eta
T^{\pi\beta}\right].
\end{align}
This leads to the condition of hydrostatic equilibrium as
\begin{align}\label{g12}
\frac{dP}{dr}+\sigma
\frac{d\Theta_{1}^{1}}{dr}+\frac{\nu'}{2}\left(\mu
+P\right)+\frac{\sigma\nu'}{2}
\left(\Theta_{1}^{1}-\Theta_{0}^{0}\right)+\frac{2\sigma}{r}
\left(\Theta_{1}^{1} -\Theta_{2}^{2}\right)=\Omega,
\end{align}
where the term $\Omega$ on right hand side of the above equation
appears due to the non-conserved nature of $f(R,T,Q)$ theory whose
value is given in Appendix \textbf{A}. Equation \eqref{g12} may be
referred as the generalized form of Tolman-Opphenheimer-Volkoff
equation that could help to illustrate systematic changes in the
self-gravitating spherically symmetric structure. We obtain a system
of four differential equations \eqref{g8}-\eqref{g10} and
\eqref{g12} which involve non-linearity, containing seven unknown
parameters
$(\nu,\chi,\mu,P,\Theta_{0}^{0},\Theta_{1}^{1},\Theta_{2}^{2})$,
thus this system is no more definite. We use systematic method
\cite{33} to close the above system and determine the unknowns. For
the field equations \eqref{g8}-\eqref{g10}, one can define the
matter variables as
\begin{equation}\label{g13}
\bar{\mu}^{(eff)}=\mu^{(eff)}-\sigma\Theta_{0}^{0},\quad
\bar{P}_{r}^{(eff)}=P^{(eff)}+\sigma\Theta_{1}^{1},
\quad \bar{P}_{\bot}^{(eff)}=P^{(eff)}+\sigma\Theta_{2}^{2}.
\end{equation}
It is obvious from the above terms that anisotropy within
self-gravitating system is induced by the source
$\Theta^{\rho}_{\eta}$. This defines the effective parameter of
anisotropy as
\begin{equation}\label{g14}
\bar{\Delta}^{(eff)}=\bar{P}_{\bot}^{(eff)}-\bar{P}_{r}^{(eff)}
=\sigma\left(\Theta_{2}^{2}-\Theta_{1}^{1}\right).
\end{equation}
It is noticeable here that the component of anisotropy disappears
for $\sigma=0$.

\section{Gravitational Decoupling}

In this section, we use gravitational decoupling via MGD approach to
solve the system \eqref{g8}-\eqref{g10}. This method serves as a
transformation of the field equations such that the newly added
factor $\Theta_{\eta}^{\rho}$ supplies the kind of effective
equations which may cause the presence of pressure anisotropy in the
interior of stellar object. The following metric represents the
solution $(\eta,\xi,\mu,P)$ corresponding to the perfect fluid as
\begin{equation}\label{g15}
ds^2=-e^{\eta}dt^2+\frac{1}{\xi}dr^2+r^2d\theta^2+r^2\sin^2\theta
d\vartheta^2,
\end{equation}
where $\eta=\eta(r)$ and $\xi=\xi(r)=1-\frac{2m}{r}$, $m$ is the
Misner-Sharp mass of the corresponding object. By imposing the
geometrical transformations of linear form on the metric potentials,
one can determine the effects of source term $\Theta_{\eta}^{\rho}$
on isotropic models as
\begin{equation}\label{g16}
\eta\rightarrow\nu=\eta+\sigma f, \quad \xi\rightarrow e^{-\chi}=\xi+\sigma t,
\end{equation}
where the two geometric deformations $t$ and $f$ are offered to
radial and temporal components, respectively. The minimal geometric
deformations ($f=0,~t\rightarrow t^*$) in the above expression
guarantees only the effects of additional source in the radial
component while the temporal component remains preserved.
Consequently, Eq.\eqref{g16} reduces to
\begin{equation}\label{g17}
\eta\rightarrow\nu=\eta, \quad \xi\rightarrow e^{-\chi}=\xi+\sigma
t^*,
\end{equation}
where $t^*=t^*(r)$. The characteristic feature of this approach is
that the source includes the quasi-decoupled system.

To workout the complex system, we divide the field equations into
two simple systems. Using the transformations \eqref{g17} in the
system \eqref{g8}-\eqref{g10}, we obtain the first set corresponding
to $\sigma=0$ as
\begin{align}\label{g18}
&8\pi\left(\mu^{(eff)}-T_{0}^{0(D)}\right)=e^{-\chi}\left(\frac{\chi'}{r}-\frac{1}{r^2}\right)
+\frac{1}{r^2},\\\label{g19}
&8\pi\left(P^{(eff)}+T_{1}^{1(D)}\right)=e^{-\chi}\left(\frac{\nu'}{r}+\frac{1}{r^2}\right)
-\frac{1}{r^2},\\\label{g20}
&8\pi\left(P^{(eff)}+T_{2}^{2(D)}\right)=-\frac{e^{-\chi}}{4}\left[\chi'\nu'-\nu'^2-2\nu''+\frac{2\chi'}{r}-\frac{2\nu'}{r}\right],
\end{align}
whereas the second set, which contains the source
$\Theta^{\rho}_{\eta}$, becomes
\begin{align}\label{g21}
&8\pi\Theta_{0}^{0}=\frac{{t^*}'}{r}+\frac{t^*}{r^2},\\\label{g22}
&8\pi\Theta_{1}^{1}=t^*\left(\frac{\nu'}{r}+\frac{1}{r^2}\right),\\\label{g23}
&8\pi\Theta_{2}^{2}=\frac{t^*}{4}\left[2\nu''+\nu'^2-\nu'\chi'+\frac{2\nu'}{r}-\frac{2\chi'}{r}\right].
\end{align}
The system \eqref{g21}-\eqref{g23} is analogous to the spherical
stellar object having anisotropy with material variables
$\bar{\mu}^{(eff)}=\Theta_{0}^{0},~\bar{P}_{r}^{(eff)}=-\Theta_{1}^{1},~\bar{P}_{\bot}^{(eff)}=-\Theta_{2}^{2}$
express the geometry
\begin{equation}\label{g24}
ds^2=-e^\nu dt^2+\frac{1}{t^*}dr^2+r^2d\theta^2+r^2\sin^2\theta
d\vartheta^2.
\end{equation}
However, Eqs.\eqref{g21}-\eqref{g23} are not typical field equations
for anisotropic spherical source as they differ by a single term
$\frac{1}{r^2}$ and thus the matter components become
$\bar{\mu}^{(eff)}=\Theta_{0}^{*0}=\Theta_{0}^{0}+\frac{1}{8\pi
r^2},\quad
\bar{P}_{r}^{(eff)}=\Theta_{1}^{*1}=\Theta_{1}^{1}+\frac{1}{8\pi
r^2},\quad
\bar{P}_{\bot}^{(eff)}=\Theta_{2}^{*2}=\Theta_{2}^{2}=\Theta_{3}^{*3}=\Theta_{3}^{3}.$
The MGD technique has therefore converted the complex system
\eqref{g8}-\eqref{g10} into a set of equations describing the
isotropic fluid ($\mu^{(eff)},P^{(eff)},\nu,\chi$) along with four
unknowns ($t^*,\Theta_{0}^{0},\Theta_{1}^{1},\Theta_{2}^{2}$)
obeying the above anisotropic system. As a result, we have decoupled
the system \eqref{g8}-\eqref{g10} successfully.

The junction conditions are very significant to examine the stellar
bodies. One can determine the fundamental characteristics of a star
via smooth matching of the exterior and interior regions. In this
case, MGD achieves the interior geometry expressed with the help of
following metric as
\begin{equation}\label{g25}
ds^2=-e^{\nu}dt^2+\frac{1}{\left(1-\frac{2\tilde{m}(r)}{r}\right)}dr^2+
r^2d\theta^2+r^2\sin^2\theta d\vartheta^2,
\end{equation}
where the interior mass is $\tilde{m}(r)=m(r)-\frac{\sigma
r}{2}t^*(r)$. To match the inner and outer sectors of a compact star
smoothly, we take the general outer metric as
\begin{equation}\label{g26}
ds^2=-e^{\nu} dt^2+e^{\chi} dr^2+r^2d\theta^2+r^2\sin^2\theta
d\vartheta^2.
\end{equation}
There are two fundamental forms of junction conditions from which
the first one ($[ds^2]_{\Sigma}=0$, where $\Sigma$ is the
hypersurface) yields
\begin{equation}\label{g27}
\nu_{-}(\mathcal{R})=\nu_{+}(\mathcal{R}), \quad
e^{-\chi_{+}(\mathcal{R})}=1-\frac{2M_{0}}{\mathcal{R}}+ \sigma
t^*(\mathcal{R}),
\end{equation}
where we have used $\xi=e^{-\chi}-\sigma t^*$. The plus and minus
signs represent outer and inner geometries, respectively. Also,
$t^*(\mathcal{R})$ and $M_{0}=m(\mathcal{R})$ represent the
deformation and total mass at the boundary $r=\mathcal{R}$. Further,
the second form ($[T_{\rho\eta}w^{\eta}]_{\Sigma}=0$) gives
\begin{equation}\label{g28}
P^{(eff)}(\mathcal{R})+\sigma
\left(\Theta^{1}_{1}(\mathcal{R})\right)_{-}+
\left(T^{1(D)}_{1}(\mathcal{R})\right)_{-}=\sigma
\left(\Theta^{1}_{1}(\mathcal{R})\right)_{+}+\left(T^{1(D)}_{1}(\mathcal{R})\right)_{+}.
\end{equation}
Using Eq.\eqref{g27}, the above equation becomes
\begin{equation}\label{g29}
P^{(eff)}(\mathcal{R})+\sigma
\left(\Theta^{1}_{1}(\mathcal{R})\right)_{-}=\sigma
\left(\Theta^{1}_{1}(\mathcal{R})\right)_{+},
\end{equation}
which, in return, gives
\begin{equation}\label{g30}
P^{(eff)}(\mathcal{R})+\frac{\sigma
t^*(\mathcal{R})}{8\pi}\left(\frac{\nu'}{\mathcal{R}}+
\frac{1}{\mathcal{R}^2}\right)=\frac{\sigma h^*(\mathcal{R})}{8\pi
\mathcal{R}^2}\left(\frac{\mathcal{R}}{\mathcal{R}-2\mathcal{M}}\right),
\end{equation}
where $\mathcal{M}$ is mass of the exterior geometry and $h^*$
denotes the exterior geometric deformation in radial component for
the Schwarzschild metric in the presence of $\Theta^{\rho}_{\eta}$
(source) given by
\begin{equation}\label{g31}
ds^2=-\left(1-\frac{2\mathcal{M}}{r}\right)dt^2+\frac{1}{\left(1-\frac{2\mathcal{M}}{r}+\sigma
h^*\right)}dr^2+ r^2d\theta^2+r^2\sin^2\theta d\vartheta^2.
\end{equation}
The two equations \eqref{g27} and \eqref{g30} provide the
appropriate and adequate conditions for discussing the relationship
between the MGD inner and outer Schwarzschild spacetimes included by
$\Theta^{\rho}_{\eta}$. One may take the usual Schwarzschild
solution, (i.e., $h^*=0$) as outer geometry, then Eq.\eqref{g30}
yields
\begin{equation}\label{g32}
\bar{P}^{(eff)}(\mathcal{R})\equiv
P^{(eff)}(\mathcal{R})+\frac{\sigma t^*(\mathcal{R})}
{8\pi}\left(\frac{1}{\mathcal{R}^2}+\frac{\nu'}{\mathcal{R}}\right)=0.
\end{equation}

\section{Anisotropic Solutions}

We take isotropic spherical solution in modified scenario to solve
the field equations corresponding to anisotropic matter configuraton
by means of MGD approach. In order to continue our analysis, we take
the Krori-Barua solution \cite{42} whose nature is non-singular.
This solution was originally developed in GR to discuss the
evolution of compact stars, but now we utilize it to construct
solutions in modified theory which produce much complicated
effective physical quantities. In $f(R,T,Q)$ framework, the solution
takes the form
\begin{eqnarray}\label{g33}
e^{\nu}&=&e^{\mathcal{B}r^2+\mathcal{C}}, \\\label{g34}
e^{\chi}&=&\xi^{-1}=e^{\mathcal{A}r^2}, \\\label{g35}
\mu^{(eff)}&=&-\frac{1}{8\pi r^2}\left[e^{-\mathcal{A}r^2}
\left(1-2\mathcal{A}r^2\right)-1\right]+T^{0(D)}_{0}, \\\label{g36}
P^{(eff)}&=&\frac{1}{8\pi
r^2}\left[e^{-\mathcal{A}r^2}\left(1+2\mathcal{B}r^2\right)-1\right]-T^{1(D)}_{1},
\end{eqnarray}
where the unknowns $\mathcal{A},~\mathcal{B}$ and $\mathcal{C}$ can
be calculated by means of smooth matching. The continuity of
$g_{tt},~g_{rr}$ and $g_{tt,r}$ (metric components) between the
inner and outer regions takes the form
\begin{eqnarray}\label{36a}
g_{tt}&=&e^{\mathcal{B}\mathcal{R}^2+C}=1-\frac{2M_0}{\mathcal{R}},\\\label{36b}
g_{rr}&=&e^{-\mathcal{A}\mathcal{R}^2}=1-\frac{2M_0}{\mathcal{R}},\\\label{36c}
\frac{\partial g_{tt}}{\partial
r}&=&\mathcal{B}\mathcal{R}e^{\mathcal{B}\mathcal{R}^2+\mathcal{C}}=\frac{M_0}{\mathcal{R}^2},
\end{eqnarray}
which after solving simultaneously leads to
\begin{eqnarray}\label{g37}
\mathcal{A}&=&\frac{1}{\mathcal{R}^2}
\ln\left(\frac{\mathcal{R}}{\mathcal{R}-2M_{0}}\right), \quad
\mathcal{B}=\frac{M_{0}}{\mathcal{R}^3}
\left(1-\frac{2M_{0}}{\mathcal{R}}\right)^{-1},\\\label{g38}
\mathcal{C}&=&\ln\left(\frac{\mathcal{R}-2M_{0}}
{\mathcal{R}}\right)-\frac{M_{0}}{\mathcal{R}}\left(1-\frac{2M_{0}}{\mathcal{R}}\right)^{-1},
\end{eqnarray}
with compactness $\frac{2M_{0}}{\mathcal{R}}<\frac{8}{9}$. At
boundary, these equations guarantee consistency of the solution
\eqref{g33}-\eqref{g36} (which we have calculated for inner
geometry) with the outer region and will be modified undoubtedly
after adding the additional source. Equations \eqref{g17} and
\eqref{g33} provide the radial and temporal metric components that
will be used for the construction of anisotropic solution, i.e., for
$\sigma\neq 0$ in the inner geometry. The relation between source
$\Theta^{\rho}_{\eta}$ and geometric deformation $t^*$ has been
expressed through Eqs.\eqref{g21}-\eqref{g23}. Further, we study a
particular compact star, namely $4U 1820-30$ with mass $M_{0}=1.58
\pm 0.06 M_{\bigodot}$ and radius $\mathcal{R}=9.1 \pm 0.4 km$
\cite{42aa}. The graphical analysis of all physical attributes is
done by using this data.

Next, we make use of some constraints to develop two feasible
solutions in the following subsections.

\subsection{Solution I}

Here, we choose a constraint depending on $\Theta_{1}^{1}$ and
calculate both $t^*$ as well as $\Theta^{\rho}_{\eta}$ to obtain the
required solution. Equation \eqref{g32} points out the compatibility
of Schwarzschild exterior geometry with interior spacetime as long
as $P^{(eff)}(\mathcal{R})+T_{1}^{1(D)}(\mathcal{R})\sim \sigma
\left(\Theta_{1}^{1}(\mathcal{R})\right)_{-}$. The easiest choice is
\cite{33}
\begin{equation}\label{g39}
P^{(eff)}+T_{1}^{1(D)}=\Theta_{1}^{1} \quad\Rightarrow\quad
t^*=\xi-\frac{1}{1+\nu'r},
\end{equation}
where we have used Eqs.\eqref{g19} and \eqref{g22}. Using this
equation, we obtain
\begin{equation}\label{g40}
e^{-\chi}=(1+\sigma)\xi-\frac{\sigma}{1+2\mathcal{B}r^2}.
\end{equation}
The two equations \eqref{g33} and \eqref{g40} contain the metric
components which characterize the Krori-Barua solution minimally
deformed by $\Theta^{\rho}_{\eta}$. It is necessary to stress that
the standard isotropic solutions \eqref{g33}-\eqref{g36} can be
found by taking $\sigma \rightarrow 0$. The continuity of the first
fundamental form gives
\begin{equation}\label{g41}
\mathcal{R}e^{\mathcal{B}\mathcal{R}^2+\mathcal{C}}=\mathcal{R}-2\mathcal{M},
\end{equation}
and
\begin{equation}\label{g42}
(1+\sigma)\xi-\frac{\sigma}{1+2\mathcal{B}\mathcal{R}^2}=1-\frac{2\mathcal{M}}{\mathcal{R}}.
\end{equation}
The second fundamental form
($P^{(eff)}(\mathcal{R})+T_{1}^{1(D)}(\mathcal{R})-\sigma
\left(\left(\Theta_{1}^{1}(\mathcal{R})\right)\right)_{-}=0$)
together with Eq.\eqref{g39} yields
\begin{equation}\label{g43}
P^{(eff)}(\mathcal{R})+T_{1}^{1(D)}(\mathcal{R})=0
\quad\Rightarrow\quad
\mathcal{A}=\frac{\ln\left(1+2\mathcal{B}\mathcal{R}^2\right)}{\mathcal{R}^2}.
\end{equation}
Also, Eq.\eqref{g42} leads to the Schwarzschild mass as
\begin{equation}\label{g44}
\frac{2\mathcal{M}}{\mathcal{R}}=\frac{2M_{0}}{\mathcal{R}}-\sigma\left(1-\frac{2M_{0}}{\mathcal{R}}\right)
+\frac{\sigma}{1+2\mathcal{B}\mathcal{R}^2}.
\end{equation}
Inserting this in Eq.\eqref{g41}, we have
\begin{equation}\label{g45}
e^{\mathcal{B}\mathcal{R}^2+\mathcal{C}}=\left(1+\sigma\right)\left(1-\frac{2M_{0}}{\mathcal{R}}\right)
-\frac{\sigma}{1+2\mathcal{B}\mathcal{R}^2}.
\end{equation}
This equation gives the constant $\mathcal{C}$ in terms of
$\mathcal{B}$. The system of equations \eqref{g43}-\eqref{g45}
offers necessary and sufficient limitations to do smooth matching
between inner and outer spacetimes. Hence, the anisotropic solution
for the case \eqref{g39} is constructed as
\begin{eqnarray}\nonumber
\bar{\mu}^{(eff)}&=&\frac{1}{8\pi
r^2}\left[e^{-\mathcal{A}r^2}\left(2\mathcal{A}r^2-1\right)\left(1+\sigma\right)+1\right]
+\frac{1}{8\pi r^2\left(1+2\mathcal{B}r^2\right)^2}\\\label{g46}
&\times&\left[\sigma-2\sigma\mathcal{B}r^2+8\pi
r^2\left(1+4\mathcal{B}r^2+4\mathcal{B}^2r^4\right)T_{0}^{0(D)}\right],
\\\label{g47} \bar{P}_{r}^{(eff)}&=&\frac{1}{8\pi
r^2}\left[\left(1+\sigma\right)
\left\{e^{-\mathcal{A}r^2}\left(2\mathcal{B}r^2+1\right)-1\right\}-8\pi
r^2T_{1}^{1(D)}\right], \\\nonumber
\bar{P}_{\bot}^{(eff)}&=&\frac{1}{8\pi
r^2}\left[e^{-\mathcal{A}r^2}\left\{1+2\mathcal{B}r^2\left(1+\sigma\right)+\sigma
\mathcal{B}r^4\left(\mathcal{B}-\mathcal{A}\right)-\sigma
\mathcal{A}r^2\right\}-1\right.\\\label{g48} &-&\left.8\pi
r^2T_{1}^{1(D)}\right]-\frac{\sigma}{8\pi\left(1+2\mathcal{B}r^2\right)}\left[\mathcal{B}
+\left(\mathcal{B}-\mathcal{A}\right)\left(1+\mathcal{B}r^2\right)\right],
\\\label{g49} \bar{\Delta}^{(eff)}&=&\frac{\sigma}{8\pi
r^2}\left(e^{-\mathcal{A}r^2}-\frac{1}{1+2\mathcal{B}r^2}\right)\left(\mathcal{B}^2r^4
-\mathcal{A}\mathcal{B}r^4-\mathcal{A}r^2-1\right).
\end{eqnarray}

\subsection{Solution II}

In this case, we take another constraint to obtain second
anisotropic solution. The constraint is taken as
\begin{equation}\label{g51}
\mu^{(eff)}-T_{0}^{0(D)}=\Theta_{0}^{0}.
\end{equation}
Making use of Eqs.\eqref{g18} and \eqref{g21}, we have
\begin{equation}\label{g52}
{t^*}'+\frac{t^*}{r}-\frac{1}{r}\left[e^{-\mathcal{A}r^2(2\mathcal{A}r^2-1)}+1\right]=0,
\end{equation}
which gives
\begin{equation}\label{g53}
t^*=\frac{\mathfrak{a}_{1}}{r}+e^{-\mathcal{A}r^2}-1,
\end{equation}
where $\mathfrak{a}_{1}$ is the constant of integration. The nature
of a solution at the core of star should be non-singular, thus we
take $\mathfrak{a}_{1}=0$ giving
\begin{equation}\label{g54}
t^*=e^{-\mathcal{A}r^2}-1.
\end{equation}
One can achieve the matching conditions by implementing the same
approach as for the first solution given as
\begin{eqnarray}\label{g55}
&&2\left(\mathcal{M}-M_{0}\right)+\sigma
\mathcal{R}\left(e^{-\mathcal{A}r^2}-1\right)=0,\\\label{g56}
&&\mathcal{B}\mathcal{R}^2+\mathcal{C}=\ln\left[1-\frac{2M_{0}}{\mathcal{R}}+\sigma
\mathcal{R}\left(e^{-\mathcal{A}r^2}-1\right)\right].
\end{eqnarray}
Finally, the expressions of
$\bar{\mu}^{(eff)},~\bar{P}_{r}^{(eff)},~\bar{P}_{\bot}^{(eff)}$ and
$\bar{\Delta}^{(eff)}$ are
\begin{align}\label{g57}
\bar{\mu}^{(eff)}&=\frac{1}{8\pi
r^2}\left[\left(1+\sigma\right)\left\{e^{-\mathcal{A}r^2}\left(2\mathcal{A}r^2-1\right)+1\right\}+8\pi
r^2T_{0}^{0(D)}\right],\\\label{g58}
\bar{P}_{r}^{(eff)}&=\frac{1}{8\pi
r^2}\left[\left(1+\sigma\right)\left\{e^{-\mathcal{A}r^2}\left(2\mathcal{B}r^2+1\right)-1\right\}
-2\sigma\mathcal{B}r^2-8\pi r^2T_{1}^{1(D)}\right],\\\nonumber
\bar{P}_{\bot}^{(eff)}&=\frac{1}{8\pi
r^2}\left[e^{-\mathcal{A}r^2}\left\{1+2\mathcal{B}r^2\left(1+\sigma\right)
+\sigma\left(\mathcal{B}^2r^4-\mathcal{A}\mathcal{B}r^4-\mathcal{A}r^2\right)\right\}-1\right.\\\label{g59}
&-\left.\sigma
r^2\left(2\mathcal{B}+\mathcal{B}^2r^2-\mathcal{A}\mathcal{B}r^2-\mathcal{A}\right)-8\pi
r^2T_{1}^{1(D)}\right],\\\label{g60}
\bar{\Delta}^{(eff)}&=\frac{\sigma}{8\pi
r^2}\left[r^2\left(e^{-\mathcal{A}r^2}-1\right)\left(\mathcal{B}^2r^2-\mathcal{A}\mathcal{B}r^2
-\mathcal{A}\right)-e^{-\mathcal{A}r^2}-4\mathcal{B}r^2+1\right].
\end{align}

\subsection{Physical Interpretation of the Obtained Solutions}

The mass of a sphere can be determined as
\begin{equation}\label{g63}
m(r)=4\pi\int_{0}^{\mathcal{R}}r^2 \bar{\mu}^{(eff)}dr.
\end{equation}
where the quantity $\bar{\mu}^{(eff)}$ describes the energy density
in $f(R,T,Q)$ gravity, whose value is provided in Eqs.\eqref{g46}
and \eqref{g57} in case of the solutions I and II, respectively. The
mass of anisotropic star can be obtained by solving this equation
numerically with condition at the center as $m(0)=0$. The
compactness factor $(\zeta(r))$ is another significant feature of
self-gravitating system. It is defined as the ratio of mass and
radius of a stellar structure. Buchdahl \cite{42a} found the maximum
value of $\zeta(r)$ by matching the inner static spherical spacetime
with outer Schwarzschild solution. For a stable star, this limit is
defined as $\zeta(r)=\frac{m}{\mathcal{R}}<\frac{4}{9}$, where
$m(r)=\frac{\mathcal{R}}{2}\left(1-e^{-\chi}\right)$. The redshift
$(D(r))$ of a self-gravitating body measures the increment in
wavelength of electromagnetic diffusion because of the gravitational
pull practiced by that body, which is given as
$D(r)=\frac{1}{\sqrt{1-2\zeta}}-1$. Buchdahl confined its value at
the surface of star as $D(r)<2$ for a perfect matter distribution.
However, its upper bound becomes $5.211$ for anisotropic configured
stellar bodies \cite{42b}.

The energy conditions are used to check the existence of ordinary
matter in the interior and viability of the resulting solutions.
These constraints are followed by the parameters governing the inner
region of the stellar objects which are made of ordinary matter. We
can categorize these bounds into dominant, strong, weak and null
energy conditions. The energy conditions in $f(R,T,Q)$ theory turn
out to be
\begin{eqnarray}\nonumber
&&\bar{\mu}^{(eff)} \geq 0, \quad
\bar{\mu}^{(eff)}+\bar{P}_{r}^{(eff)} \geq 0,\\\nonumber
&&\bar{\mu}^{(eff)}+\bar{P}_{\bot}^{(eff)} \geq 0, \quad
\bar{\mu}^{(eff)}-\bar{P}_{r}^{(eff)} \geq 0,\\\label{g50}
&&\bar{\mu}^{(eff)}-\bar{P}_{\bot}^{(eff)} \geq 0, \quad
\bar{\mu}^{(eff)}+\bar{P}_{r}^{(eff)}+2\bar{P}_{\bot}^{(eff)} \geq
0.
\end{eqnarray}
The stability of a stellar object is found to be a key factor in
astrophysics to analyze a feasible system. We examine stability by
taking the causality condition according to which the square of
sound speed within the geometrical structure must lie in the range
$[0,1]$, i.e., $0 \leq v_{s}^{2} < 1$. For anisotropic matter
configuration, the difference between sound speeds in tangential
$(v_{s\bot}^{2}=\frac{dP_{\bot}}{d\mu})$ and radial directions
$(v_{sr}^{2}=\frac{dP_{r}}{d\mu})$ can be used to check the stable
region of compact structures as $\mid v_{s\bot}^{2}-v_{sr}^{2} \mid
<1$ \cite{42ba}. The term $v_s^2=v_{sr}^{2}+v_{s\bot}^{2}$ also
guarantees stability of the resulting solution if it is less than
one throughout the structure. An adiabatic index $(\Gamma)$ also
plays a crucial role in analyzing the stability of compact stars.
For a stable stellar structure, the value of $\Gamma$ should not be
less than $\frac{4}{3}$ \cite{42c}-\cite{42e}. Here,
$\Gamma^{(eff)}$ can be expressed as
\begin{equation}\label{g62}
\Gamma^{(eff)}=\frac{\bar{\mu}^{(eff)}+\bar{P}_{r}^{(eff)}}{\bar{P}_{r}^{(eff)}}
\left(\frac{d\bar{P}_{r}^{(eff)}}{d\bar{\mu}^{(eff)}}\right).
\end{equation}
\begin{figure}\center
\epsfig{file=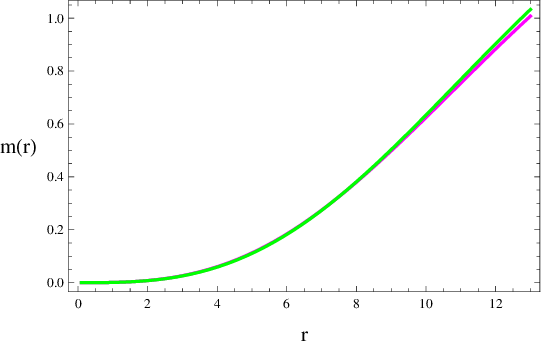,width=0.4\linewidth}
\epsfig{file=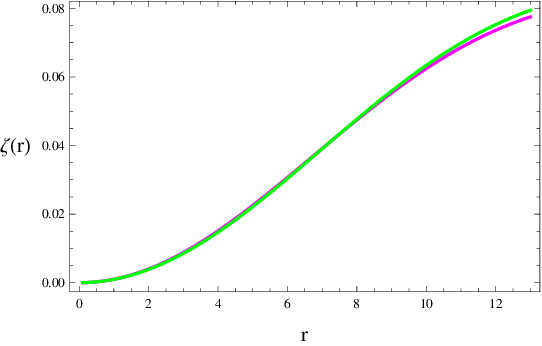,width=0.4\linewidth}
\epsfig{file=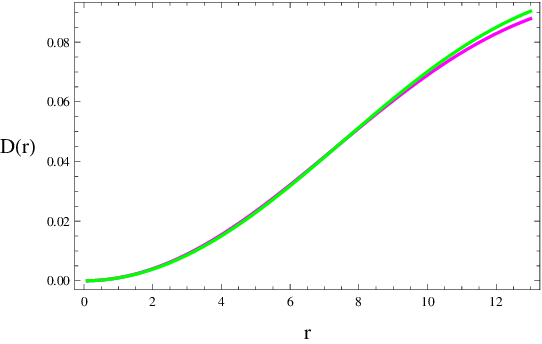,width=0.4\linewidth} \caption{Graphical
analysis of some physical parameters corresponding to $\sigma=0.1$
(pink) and $\sigma=0.9$ (green) for solution-I.}
\end{figure}

In order to discuss physical viability and stability of the obtained
solutions, we take the following model \cite{22}
\begin{equation}\label{g61}
f(R,T,R_{\rho\eta}T^{\rho\eta})=R+\alpha R_{\rho\eta}T^{\rho\eta},
\end{equation}
where $\alpha$ works as the coupling constant. Here, $\alpha$ can be
positive or negative. For its positive values, the matter variables
such as energy density and radial/tangential pressures corresponding
to both resulting solutions do not show acceptable behavior. Thus,
we are left only with negative values of $\alpha$ and we take it as
$-0.3$ to analyze physical nature of the solution-I and fix the
constant $\mathcal{A}$ calculated in Eq.\eqref{g43}. The remaining
two constants $\mathcal{B}$ and $\mathcal{C}$ are given in
Eqs.\eqref{g37} and \eqref{g38}. Figure \textbf{1} (left) shows mass
of the geometry \eqref{g6} for $\sigma=0.1$ and 0.9. It is observed
that mass increases with rise in the decoupling parameter $\sigma$.
The other two plots of Figure \textbf{1} point out that the ranges
of compactness factor and redshift parameter agree with their
respective bounds.
\begin{figure}\center
\epsfig{file=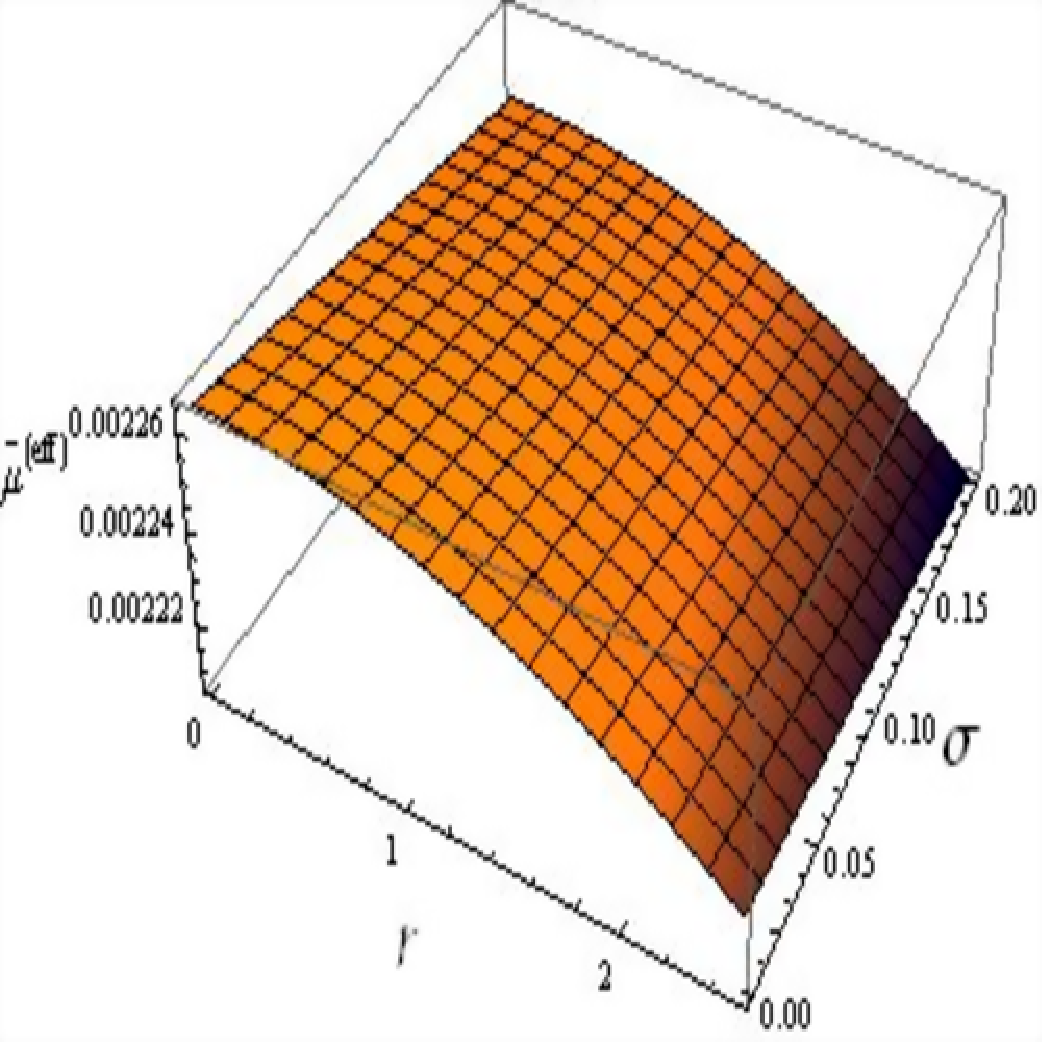,width=0.4\linewidth}\epsfig{file=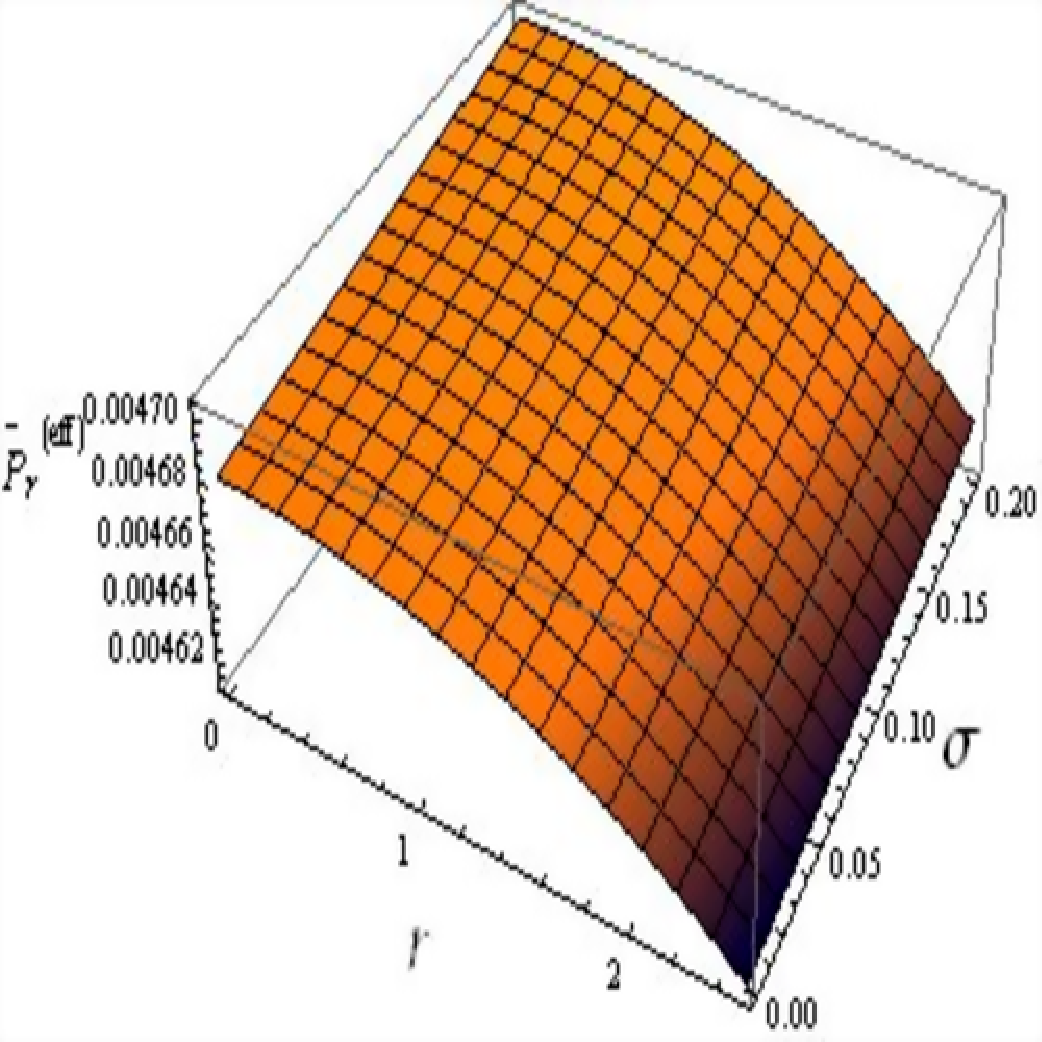,width=0.4\linewidth}
\epsfig{file=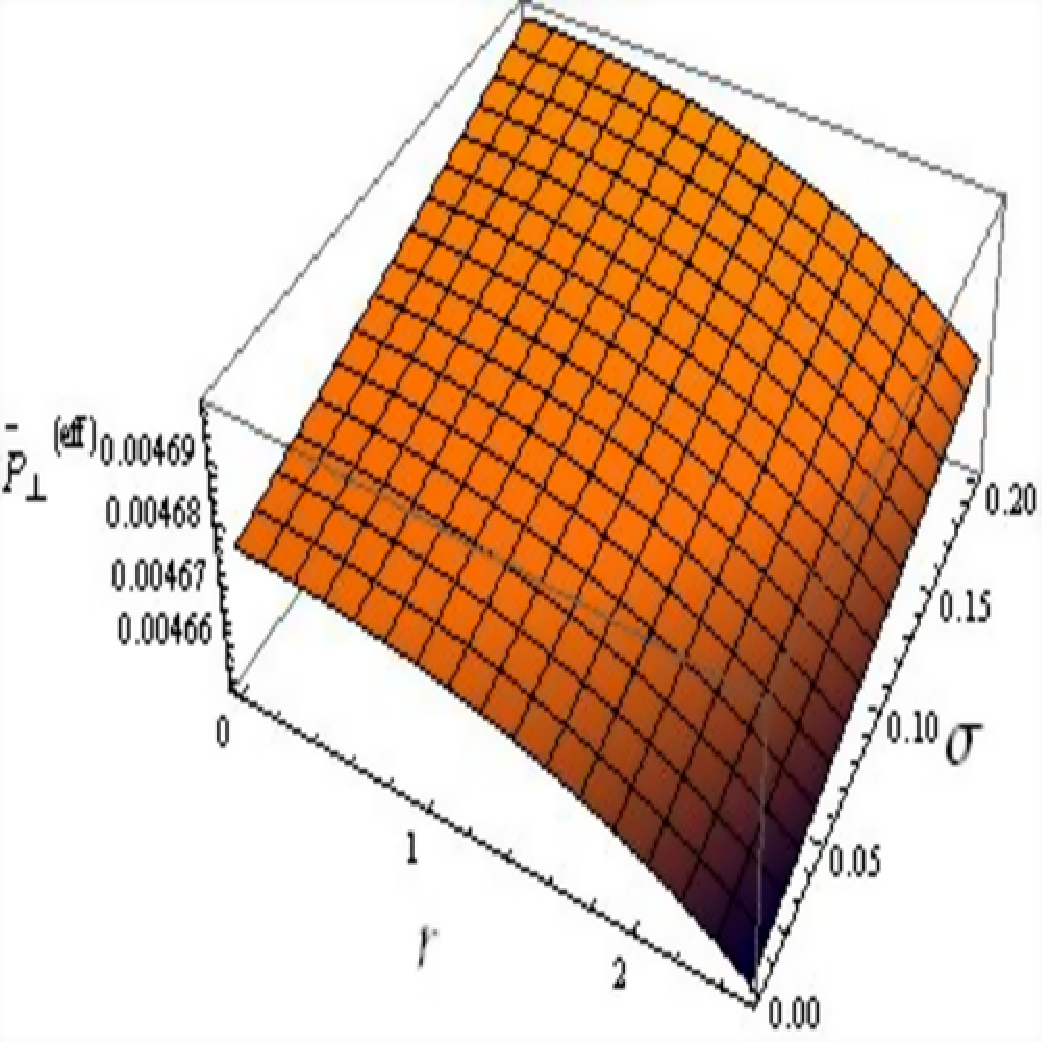,width=0.4\linewidth}\epsfig{file=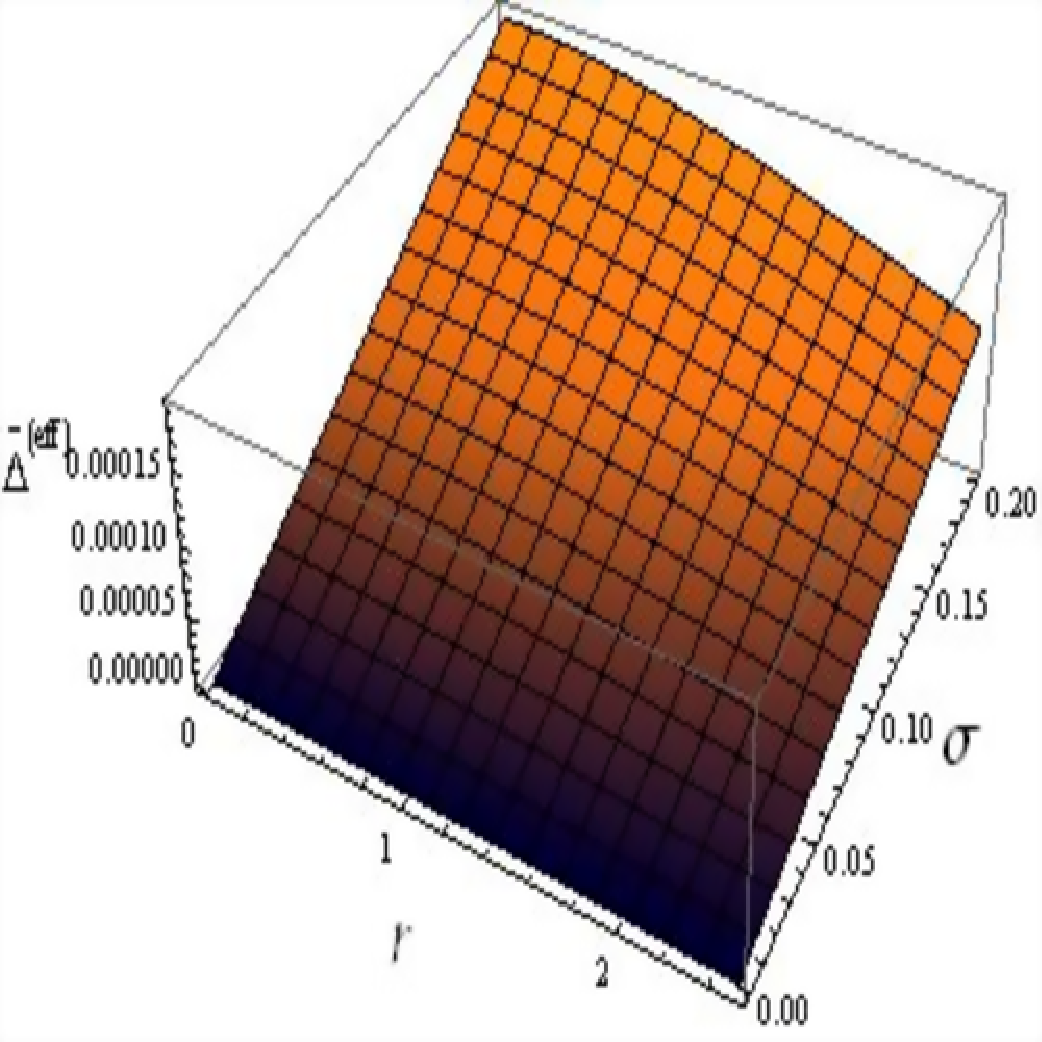,width=0.4\linewidth}
\caption{Plots of
$\bar{\mu}^{(eff)},~\bar{P}_{r}^{(eff)},~\bar{P}_{\bot}^{(eff)}$ and
$\bar{\Delta}^{(eff)}$ versus $r$ and $\sigma$ for solution-I.}
\end{figure}
\begin{figure}\center
\epsfig{file=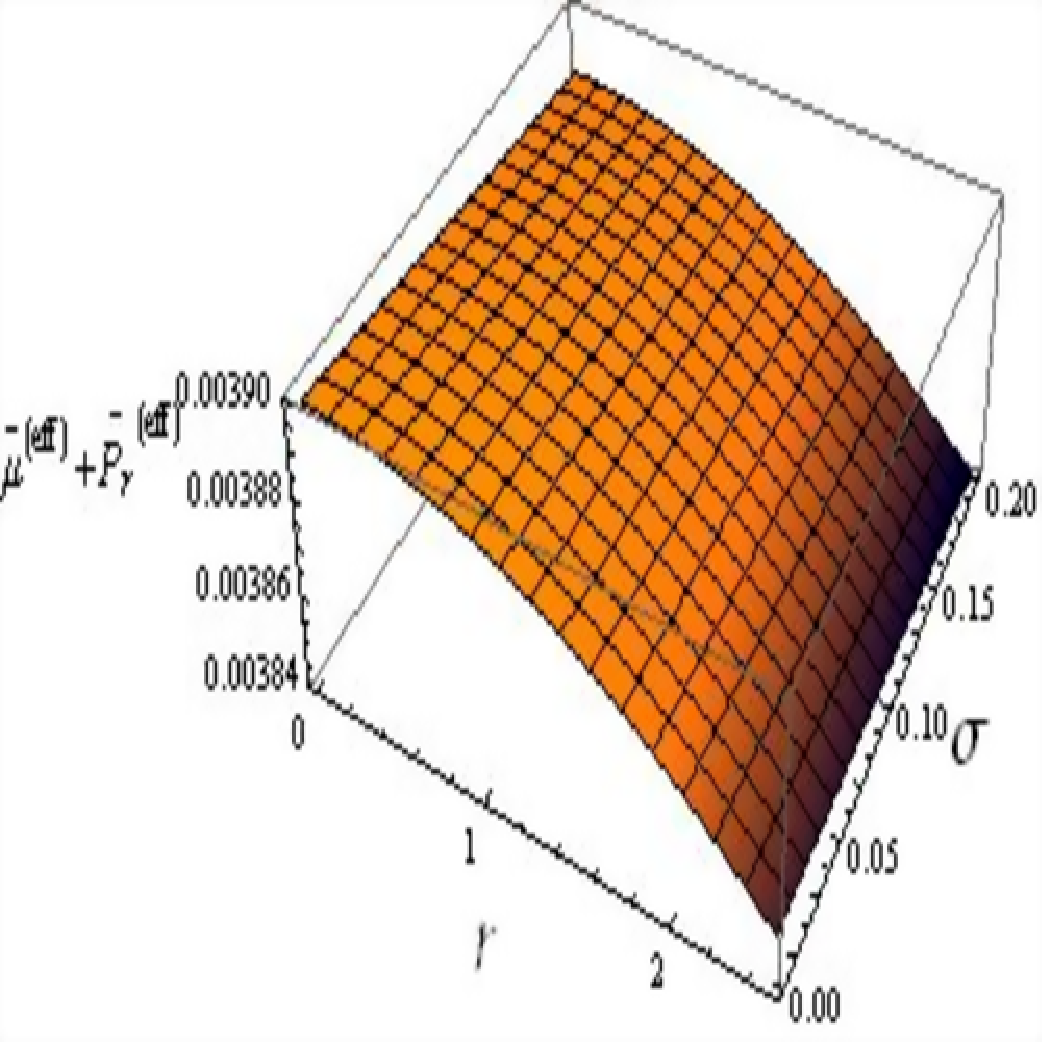,width=0.4\linewidth}\epsfig{file=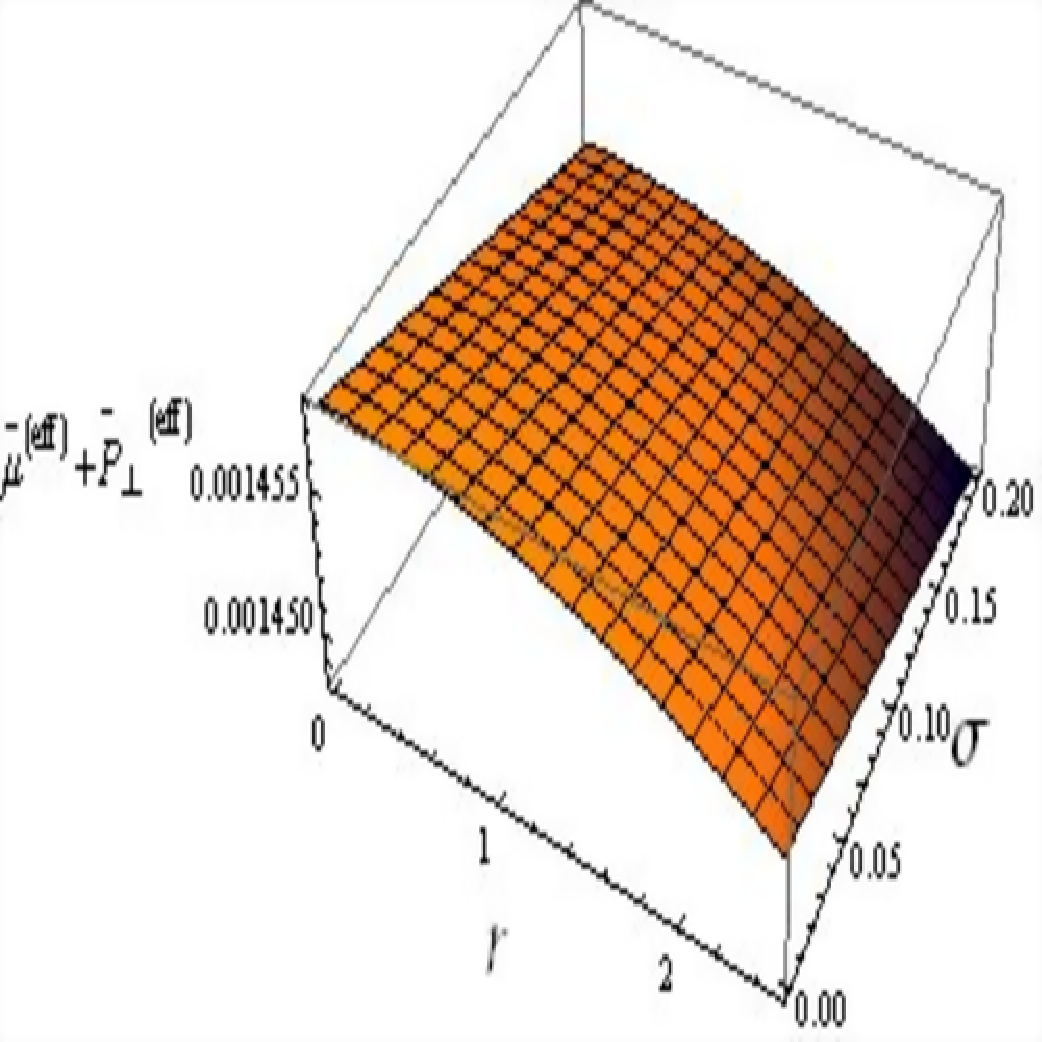,width=0.4\linewidth}
\epsfig{file=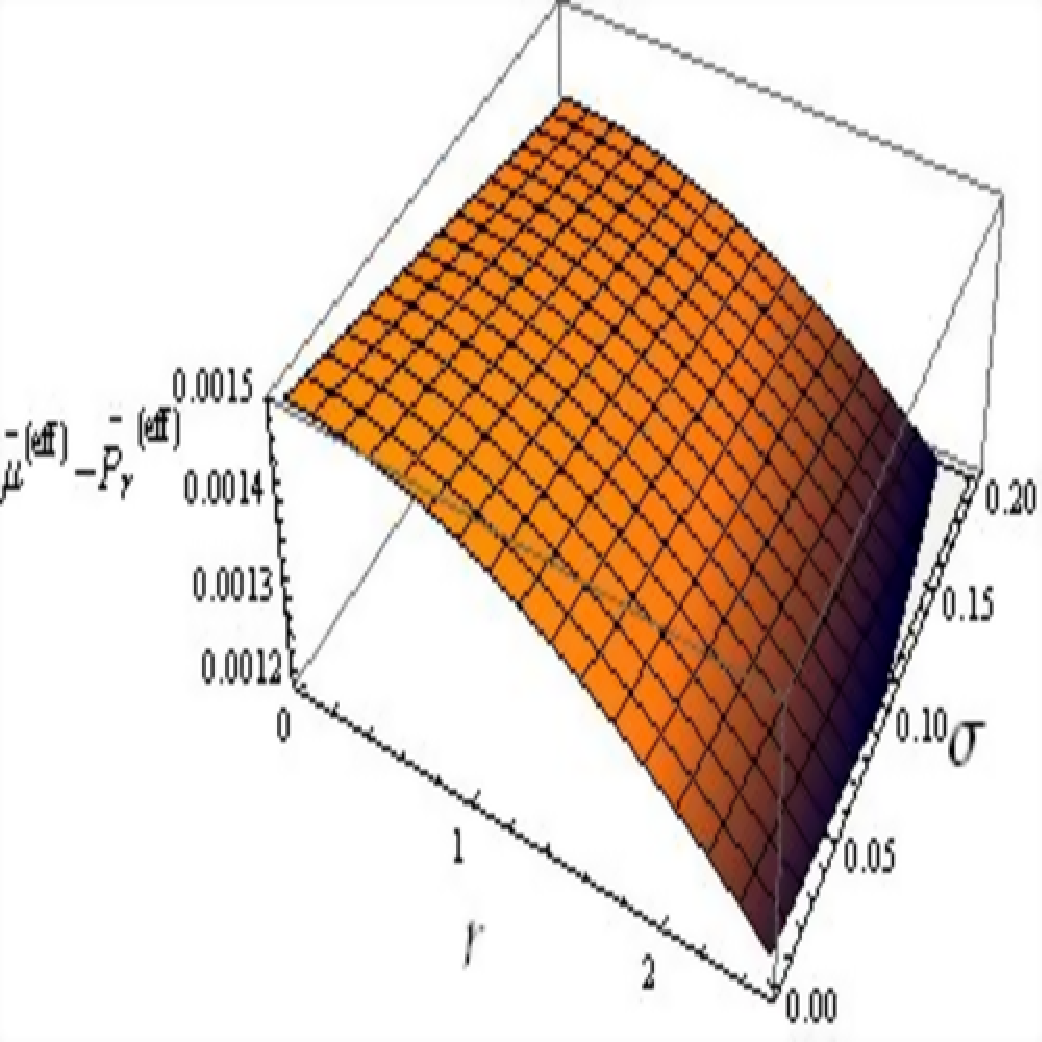,width=0.4\linewidth}\epsfig{file=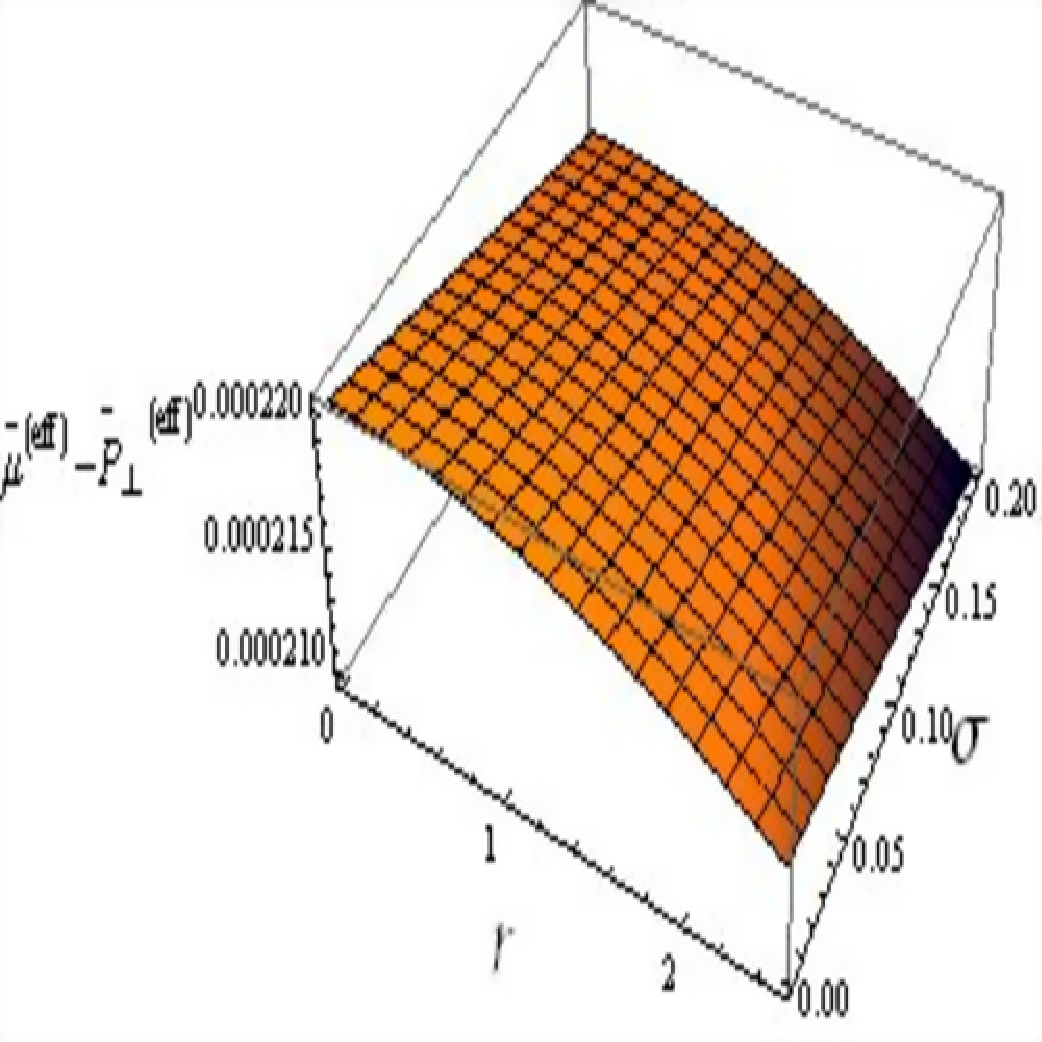,width=0.4\linewidth}
\epsfig{file=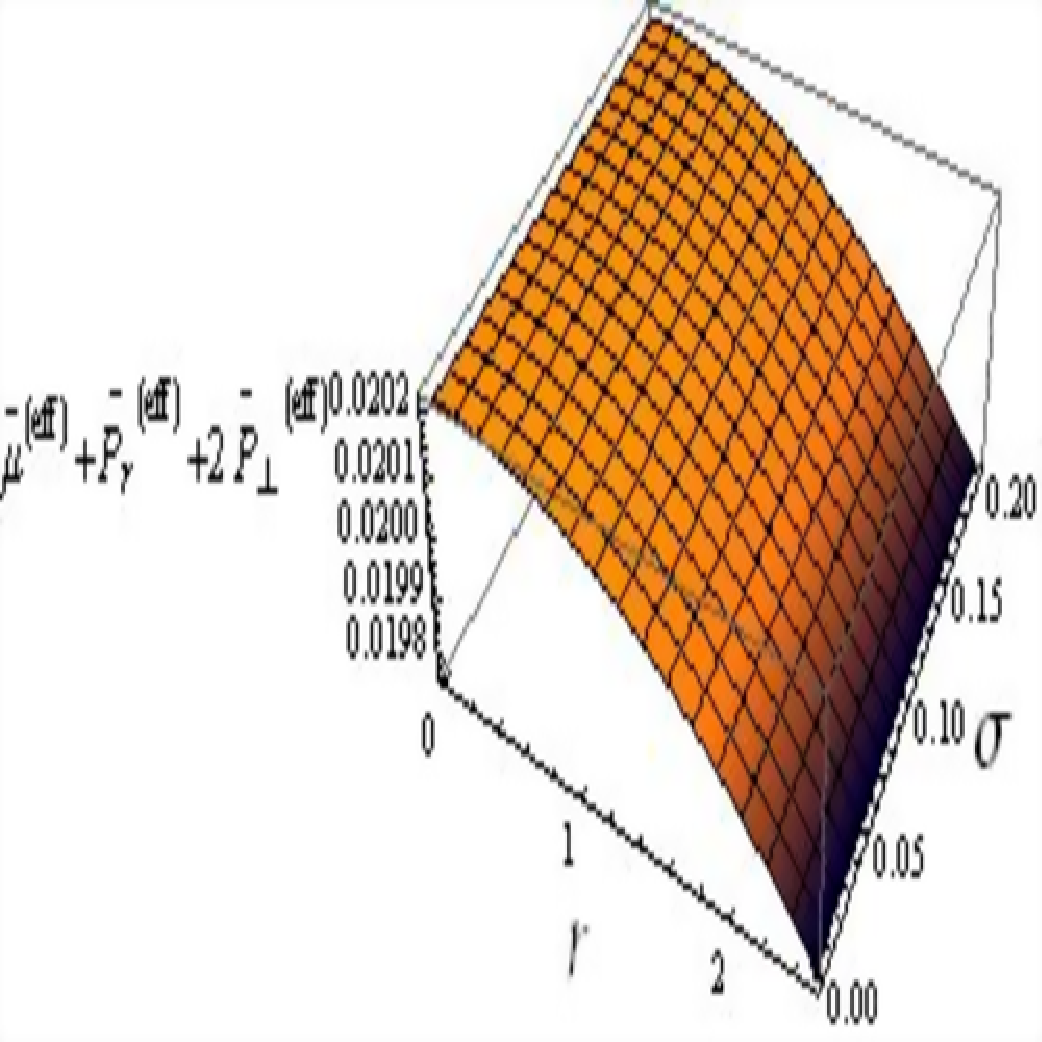,width=0.4\linewidth} \caption{Behavior of
energy bounds versus $r$ and $\sigma$ for solution-I.}
\end{figure}

For an astrophysical object, the value of material variables (such
as energy density and radial as well as tangential pressures) should
be finite, maximum and positive at the center. Further, their
behavior towards the star's boundary must be monotonically
decreasing. One can analyze from Figure \textbf{2} (upper left) that
the energy density involving $f(R,T,Q)$ corrections is maximal at
the center while shows linearly decreasing behavior towards
boundary. It is noted that the increment in $\sigma$ also decreases
energy density. The graphical nature of $\bar{P}_{r}^{(eff)}$ and
$\bar{P}_{\bot}^{(eff)}$ is shown similar to each other for the
parameter $\alpha$. By increasing the value of $r$, both ingredients
decrease as well as there is a gradual linear increment in
$\bar{P}_{\bot}^{(eff)}$ with rise in $\sigma$ as compared to
$\bar{P}_{r}^{(eff)}$. The factor $\bar{\Delta}^{(eff)}$ in Figure
\textbf{2} (lower right) shows positive behavior and increases with
the increase in the decoupling parameter $\sigma$. This indicates
that $\sigma$ generates stronger anisotropy in the structure. The
values of radial and tangential pressures are equal at the center
thus anisotropy disappears at that point. The system will be
considered viable if it meets all the energy bounds \eqref{g50}.
Figure \textbf{3} shows that our developed anisotropic solution-I is
physically viable as all energy conditions are satisfied. Figure
\textbf{4} demonstrates that the solution-I \eqref{g46}-\eqref{g49}
fulfills stability criteria for all values of the decoupling
parameter.
\begin{figure}\center
\epsfig{file=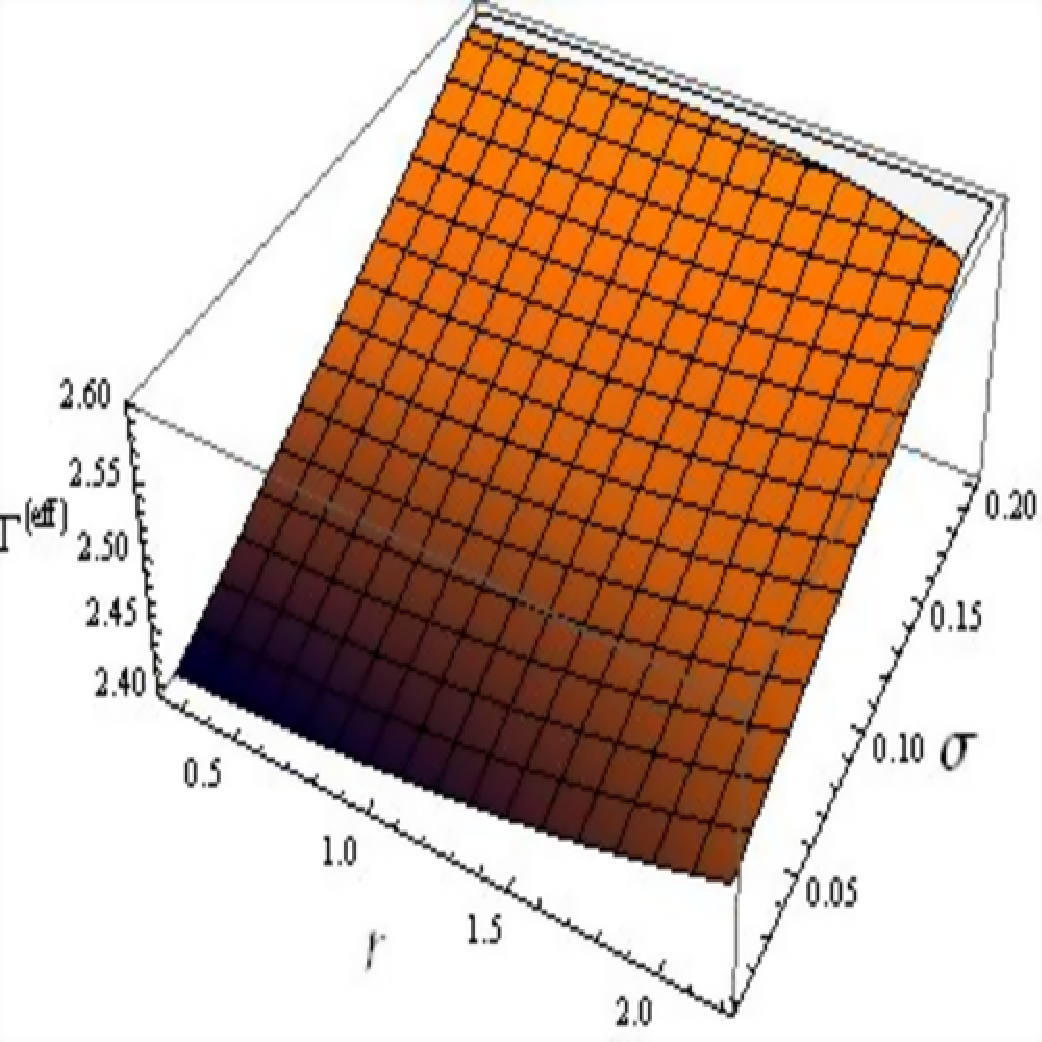,width=0.4\linewidth}\epsfig{file=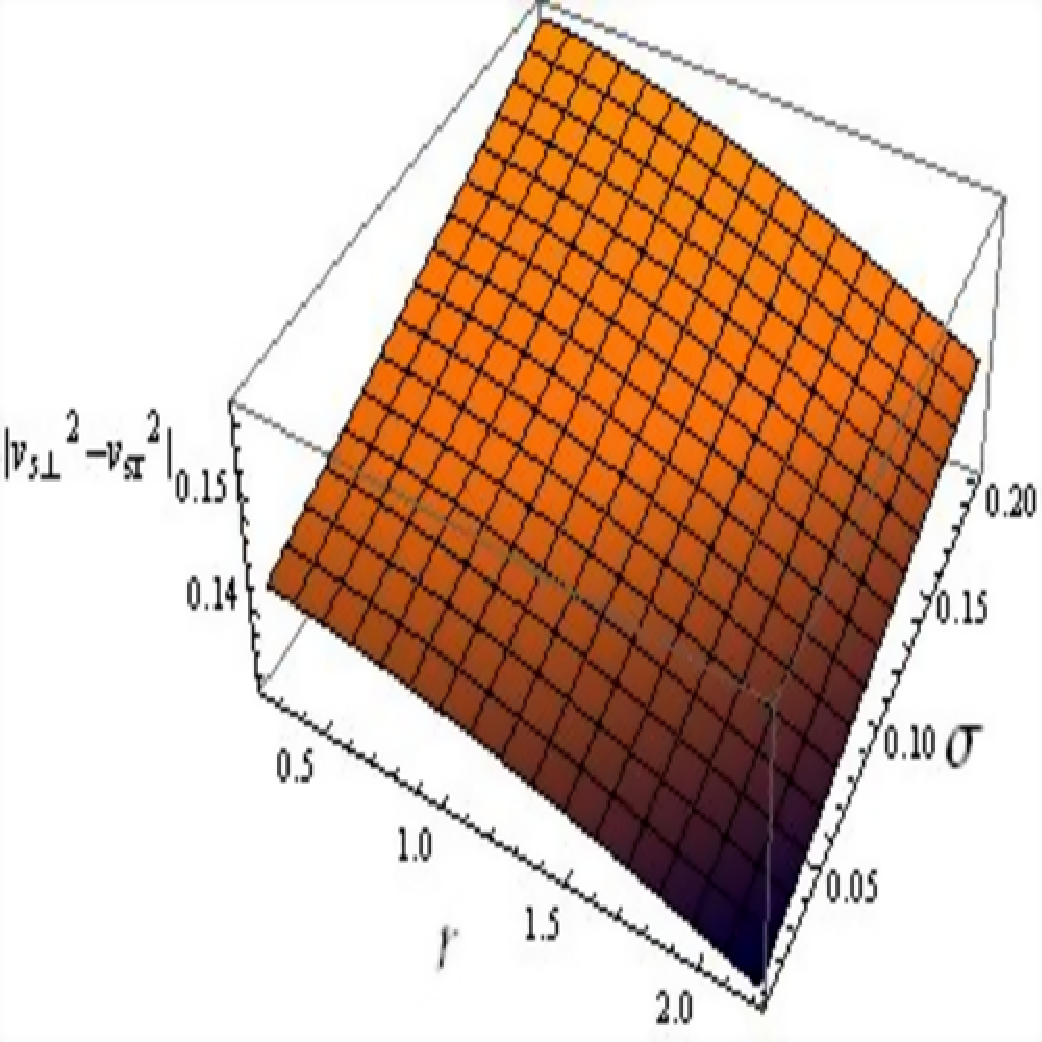,width=0.4\linewidth}
\caption{Plots of adiabatic index and $|v_{s\bot}^2-v_{sr}^2|$
versus $r$ and $\sigma$ for solution-I.}
\end{figure}

Now we explore physical features of the second solution by taking
same value of $\alpha$ as for solution-I. The constants
$\mathcal{A}$ and $\mathcal{B}$ are presented in Eqs.\eqref{g37} and
\eqref{g56}. Figure \textbf{5} (upper left) indicates that the mass
of self-gravitating body shows decreasing behavior as the parameter
$\sigma$ increases. The parameters $D(r)$ and $\zeta(r)$ also meet
the desired limits as can be seen from Figure \textbf{5}. The
physical behavior of
$\bar{\mu}^{(eff)},~\bar{P}_{r}^{(eff)},~\bar{P}_{\bot}^{(eff)}$ and
$\bar{\Delta}^{(eff)}$ is shown in Figure \textbf{6}. When one
increases the value of $\sigma$, $\bar{\mu}^{(eff)}$ and both
effective pressure components show increasing and decreasing
behavior, respectively. The lower right plot in Figure \textbf{6}
indicates that $\bar{\Delta}^{(eff)}$ shows increasing behavior with
the rise in $\sigma$ which produces stronger anisotropy in the
system. Figure \textbf{7} guarantees the regular behavior of both
solutions as
$\frac{d\bar{\mu}^{(eff)}}{dr}<0,~\frac{d\bar{P}_{r}^{(eff)}}{dr}<0$
and $\frac{\bar{P}_{\bot}^{(eff)}}{dr}<0$ everywhere. Figure
\textbf{8} shows that all energy constraints \eqref{g50} for
solution-II are satisfied and hence it is physically viable. Figure
\textbf{9} reveals that our second solution \eqref{g57}-\eqref{g60}
is also stable everywhere. Figure \textbf{10} also confirms
stability of both the developed solutions.
\begin{figure}\center
\epsfig{file=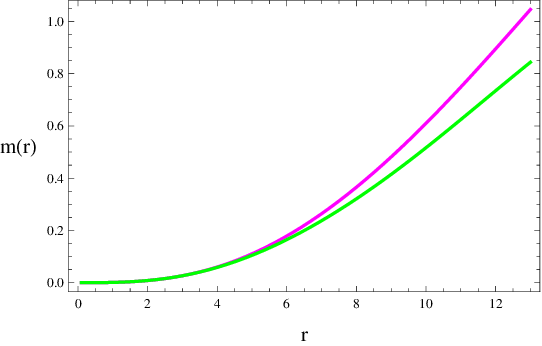,width=0.4\linewidth}
\epsfig{file=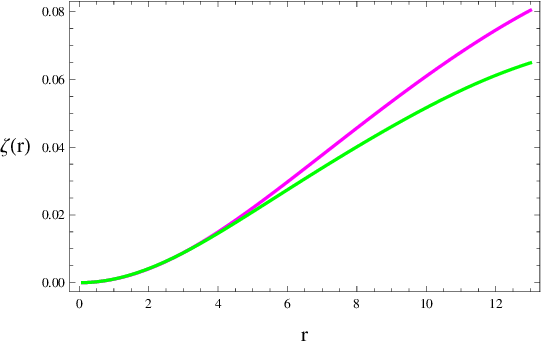,width=0.4\linewidth}
\epsfig{file=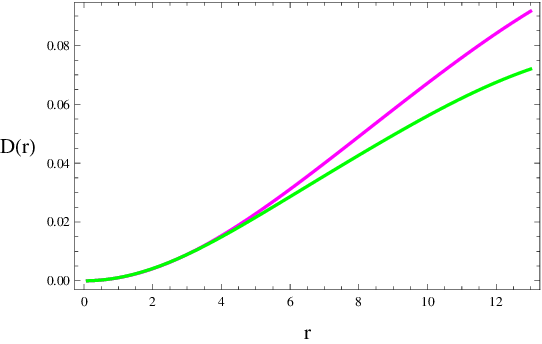,width=0.4\linewidth} \caption{Graphical
analysis of some physical parameters corresponding to $\sigma=0.1$
(pink) and $\sigma=0.3$ (green) for solution-II.}
\end{figure}
\begin{figure}\center
\epsfig{file=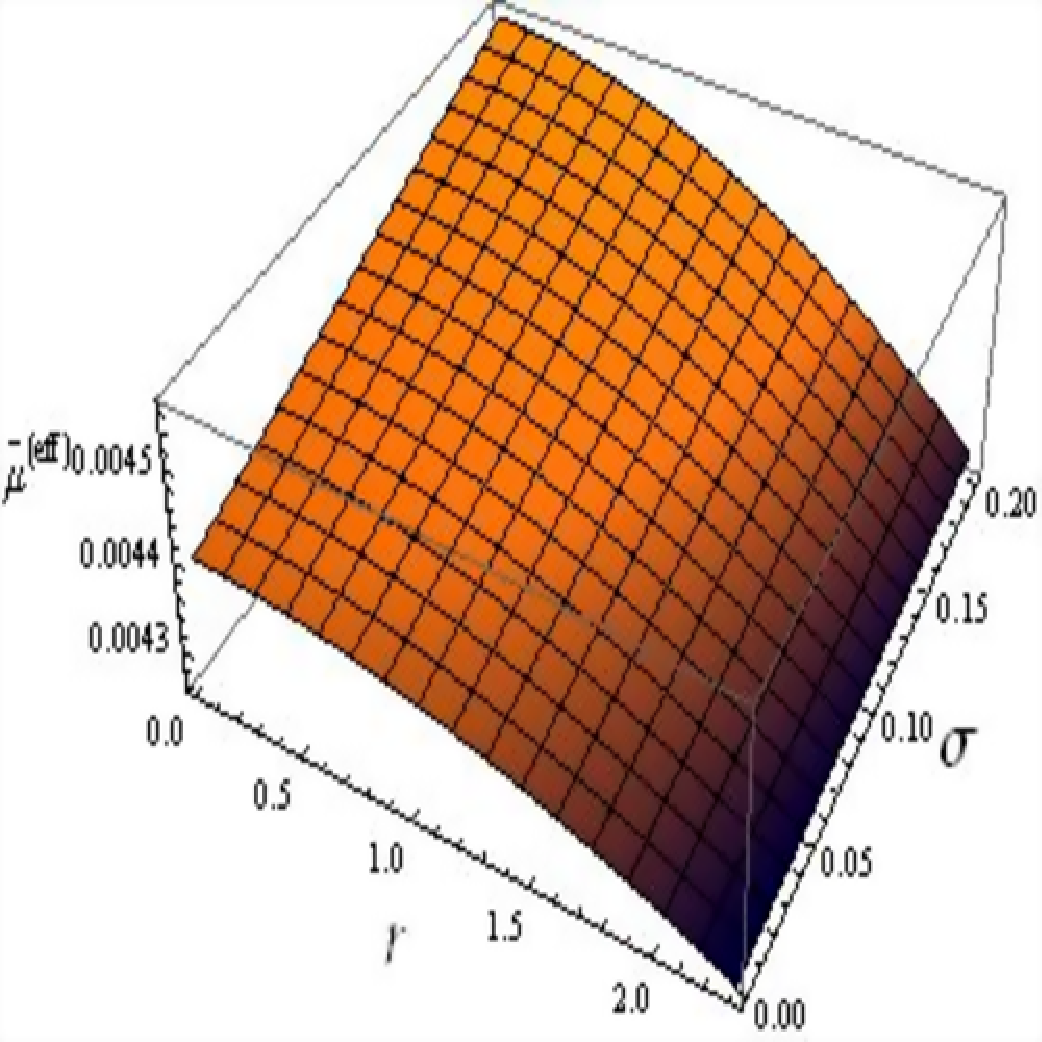,width=0.4\linewidth}\epsfig{file=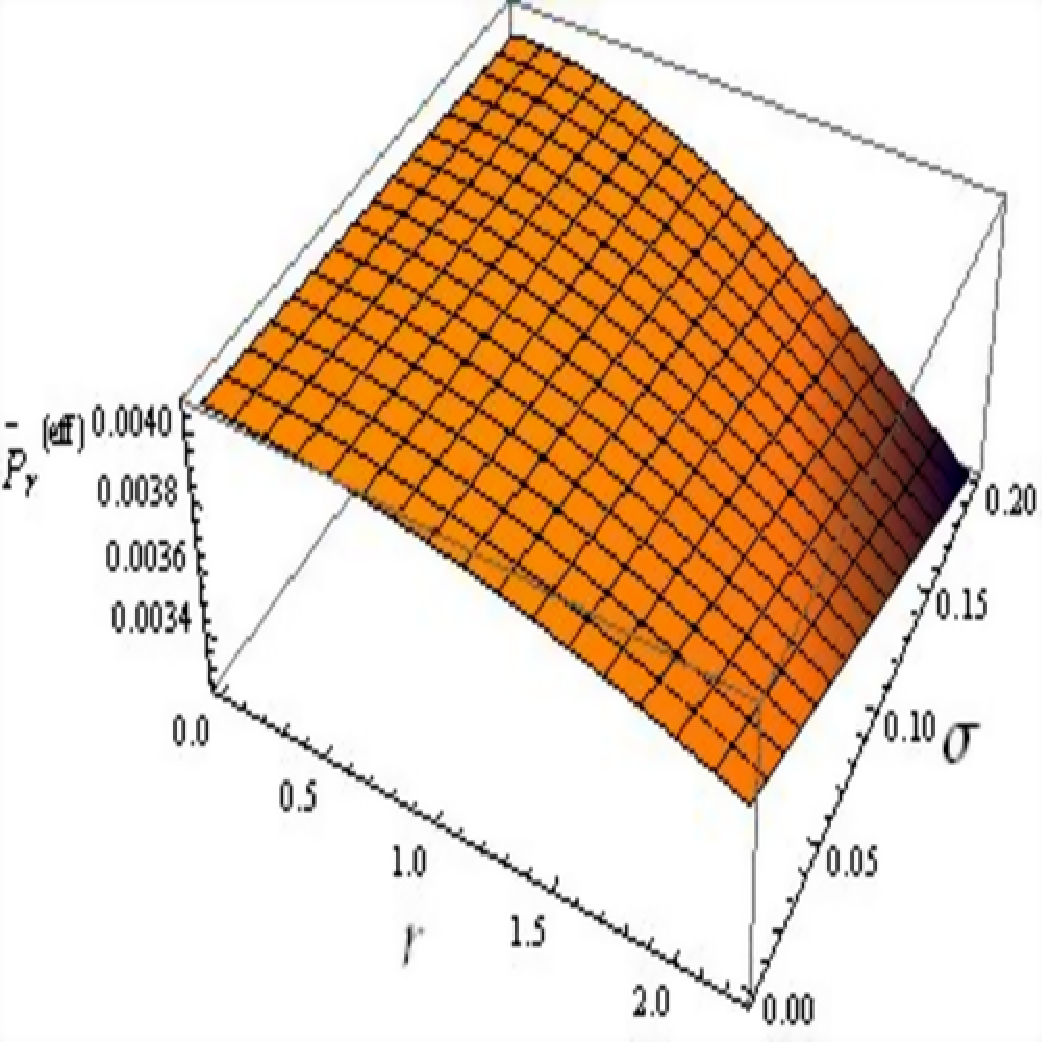,width=0.4\linewidth}
\epsfig{file=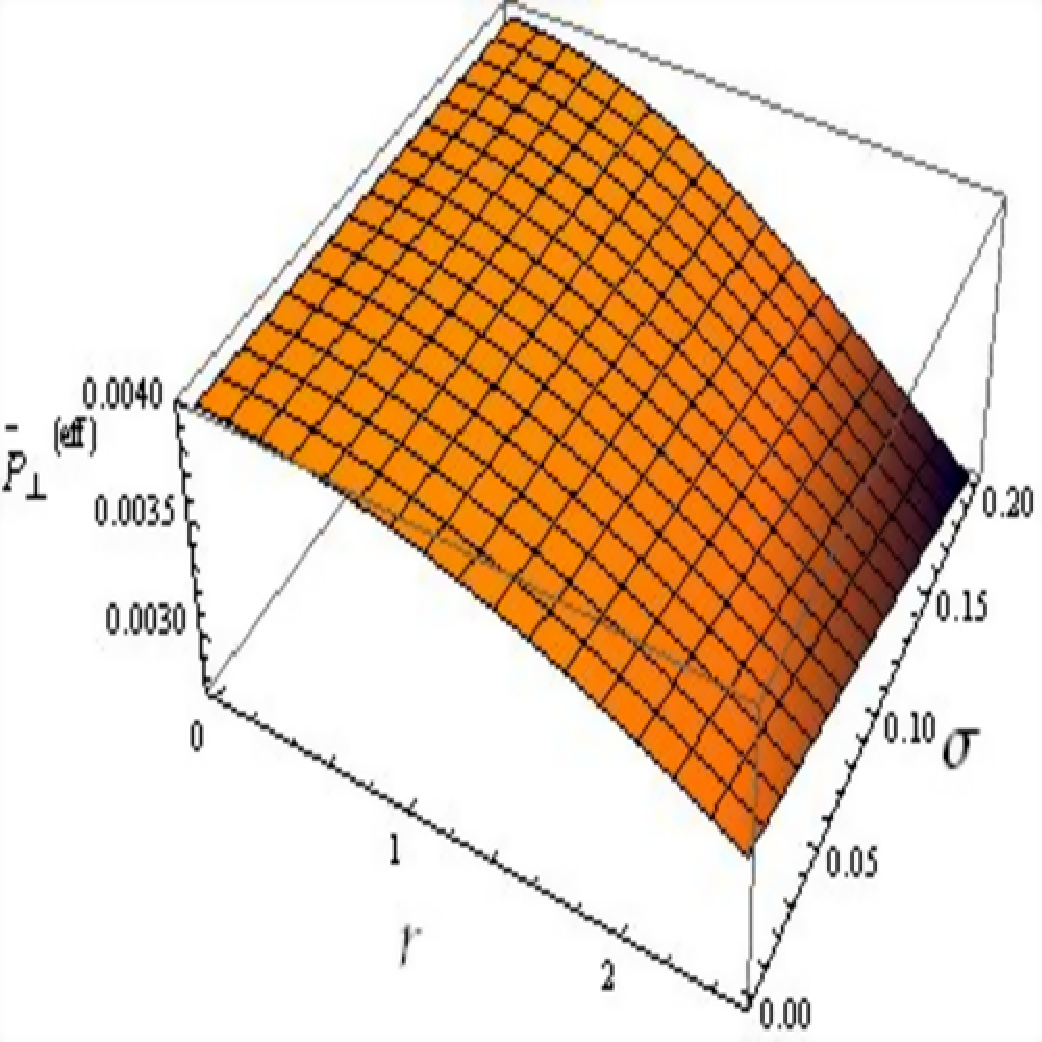,width=0.4\linewidth}\epsfig{file=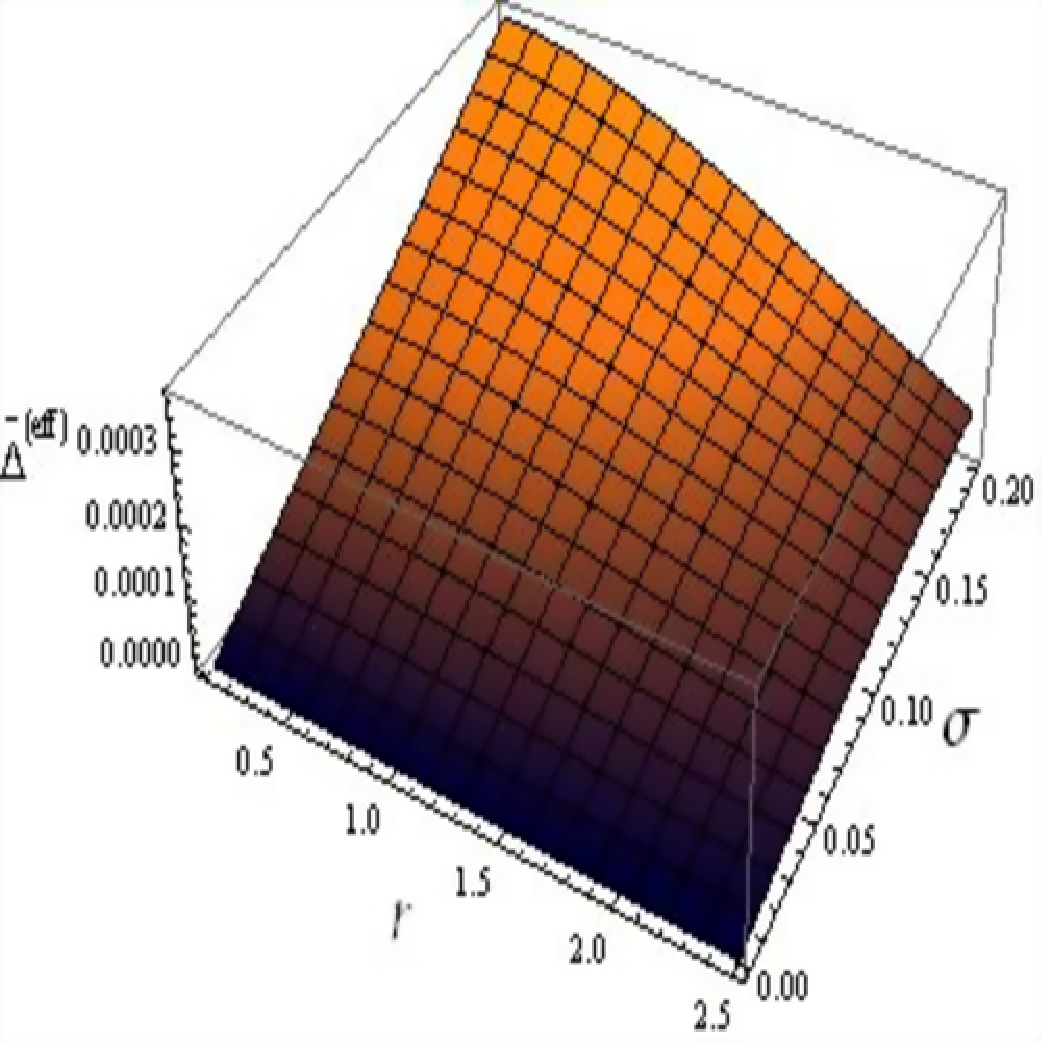,width=0.4\linewidth}
\caption{Plots of
$\bar{\mu}^{(eff)},~\bar{P}_{r}^{(eff)},~\bar{P}_{\bot}^{(eff)}$ and
$\bar{\Delta}^{(eff)}$ versus $r$ and $\sigma$ for solution-II.}
\end{figure}
\begin{figure}\center
\epsfig{file=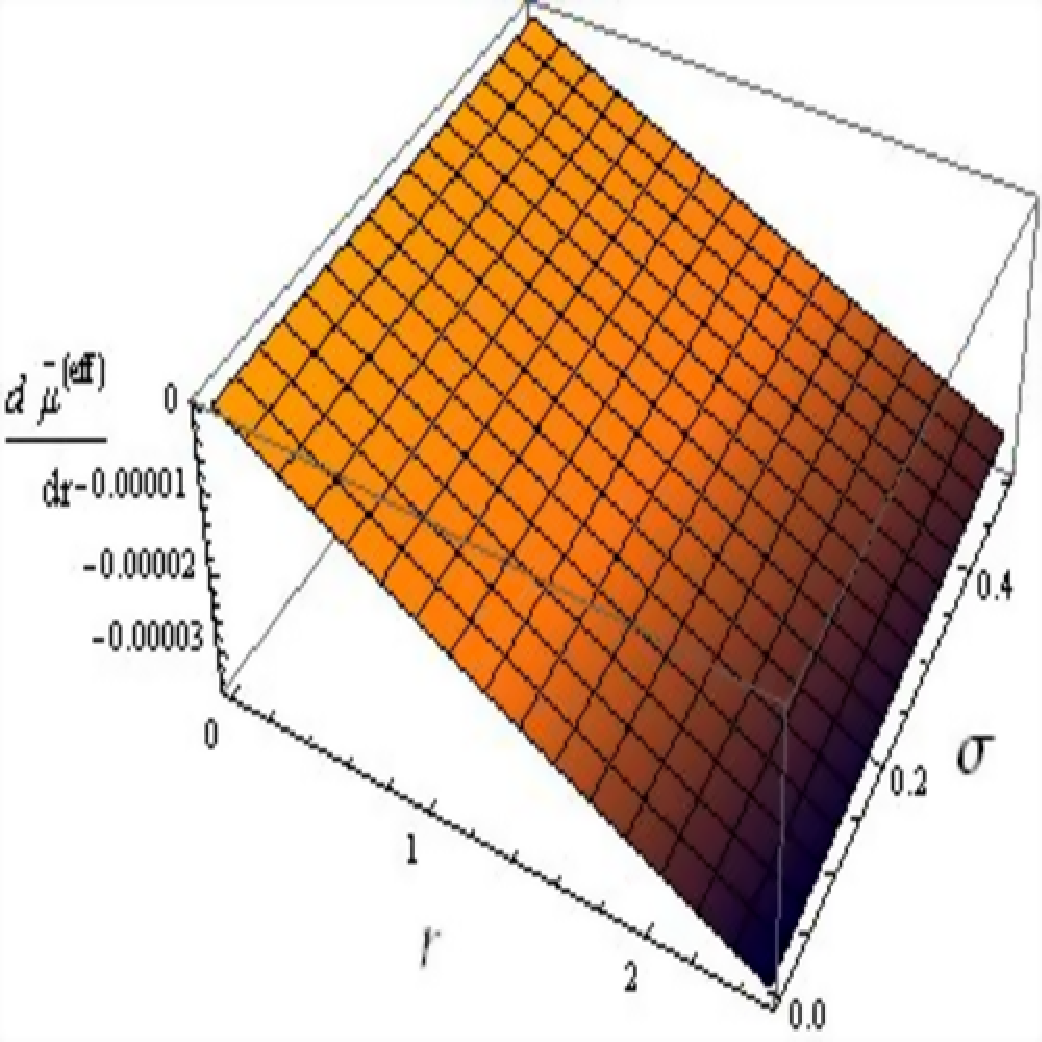,width=0.4\linewidth}\epsfig{file=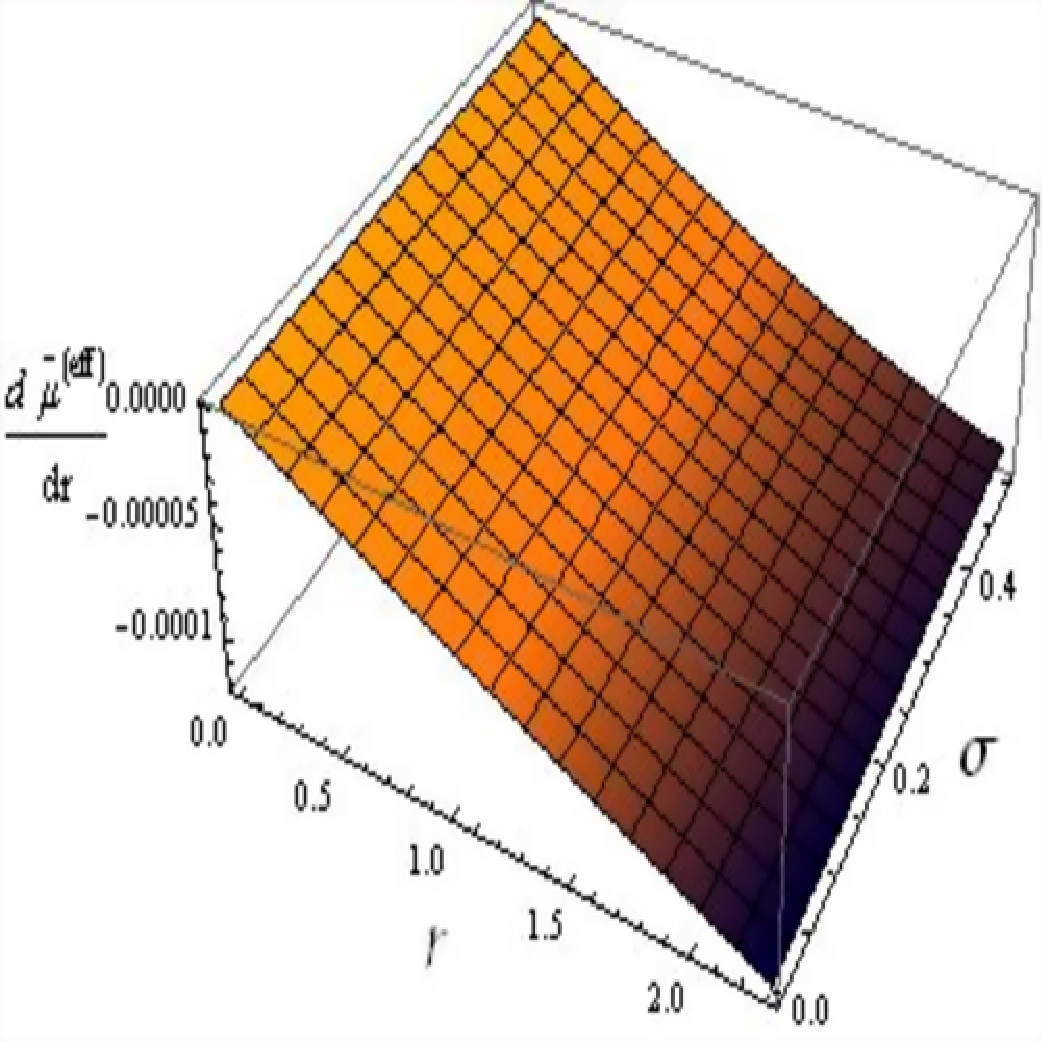,width=0.4\linewidth}
\epsfig{file=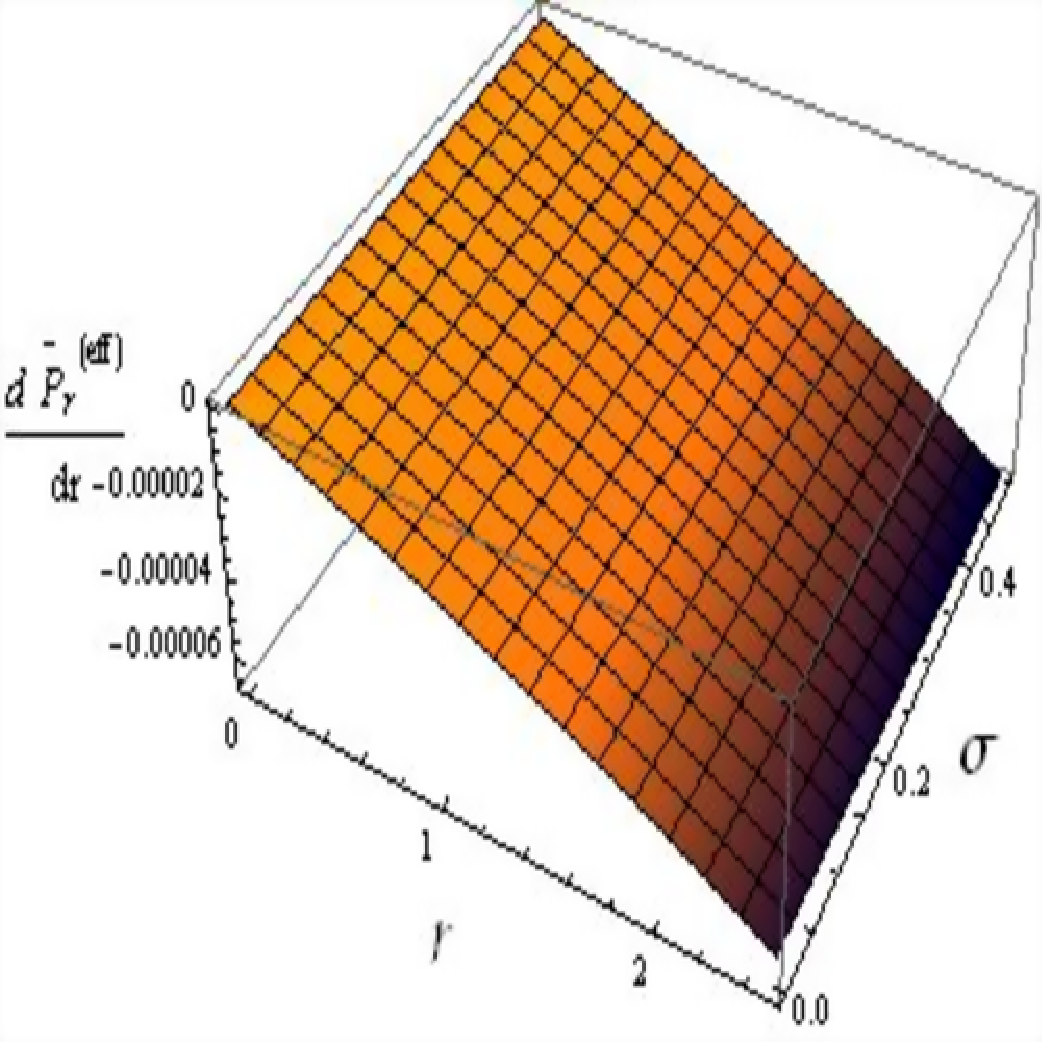,width=0.4\linewidth}\epsfig{file=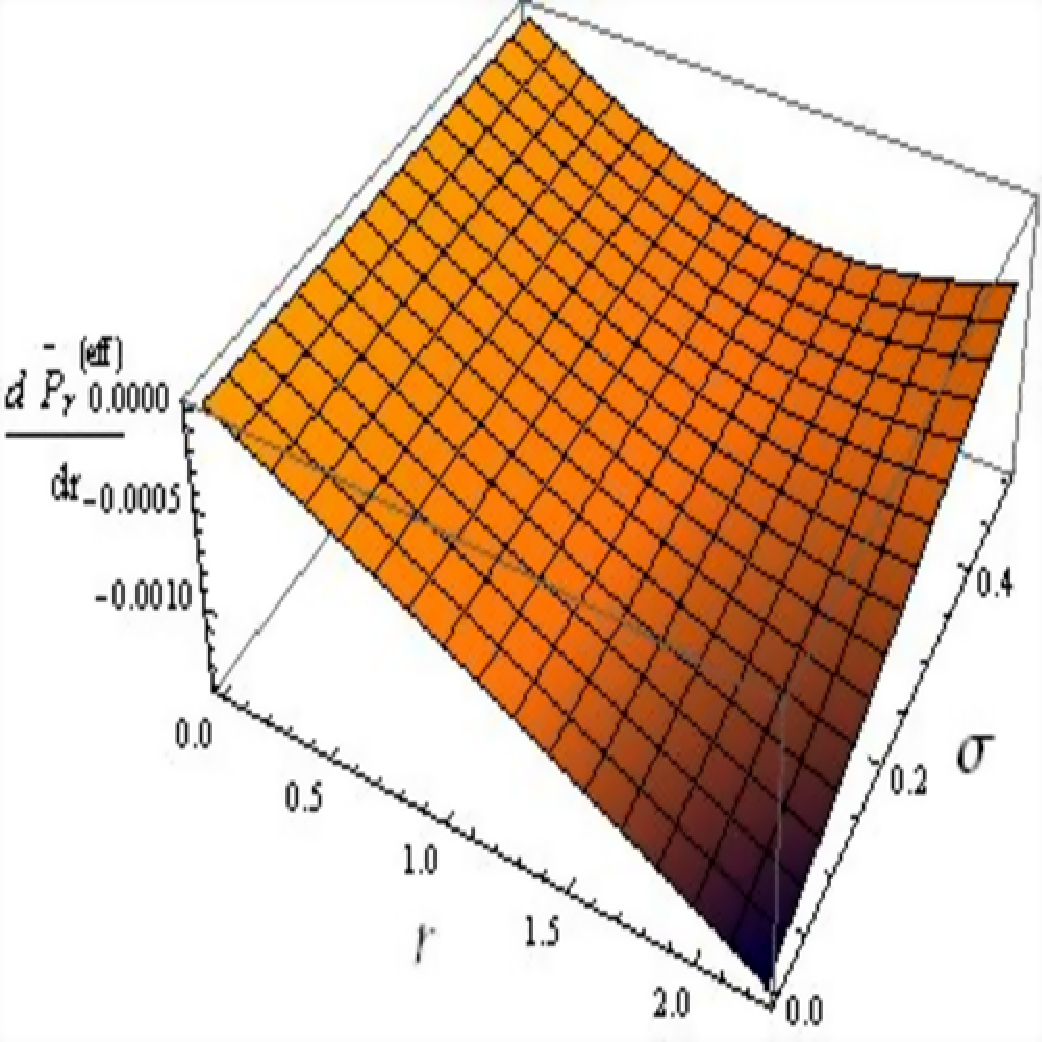,width=0.4\linewidth}
\epsfig{file=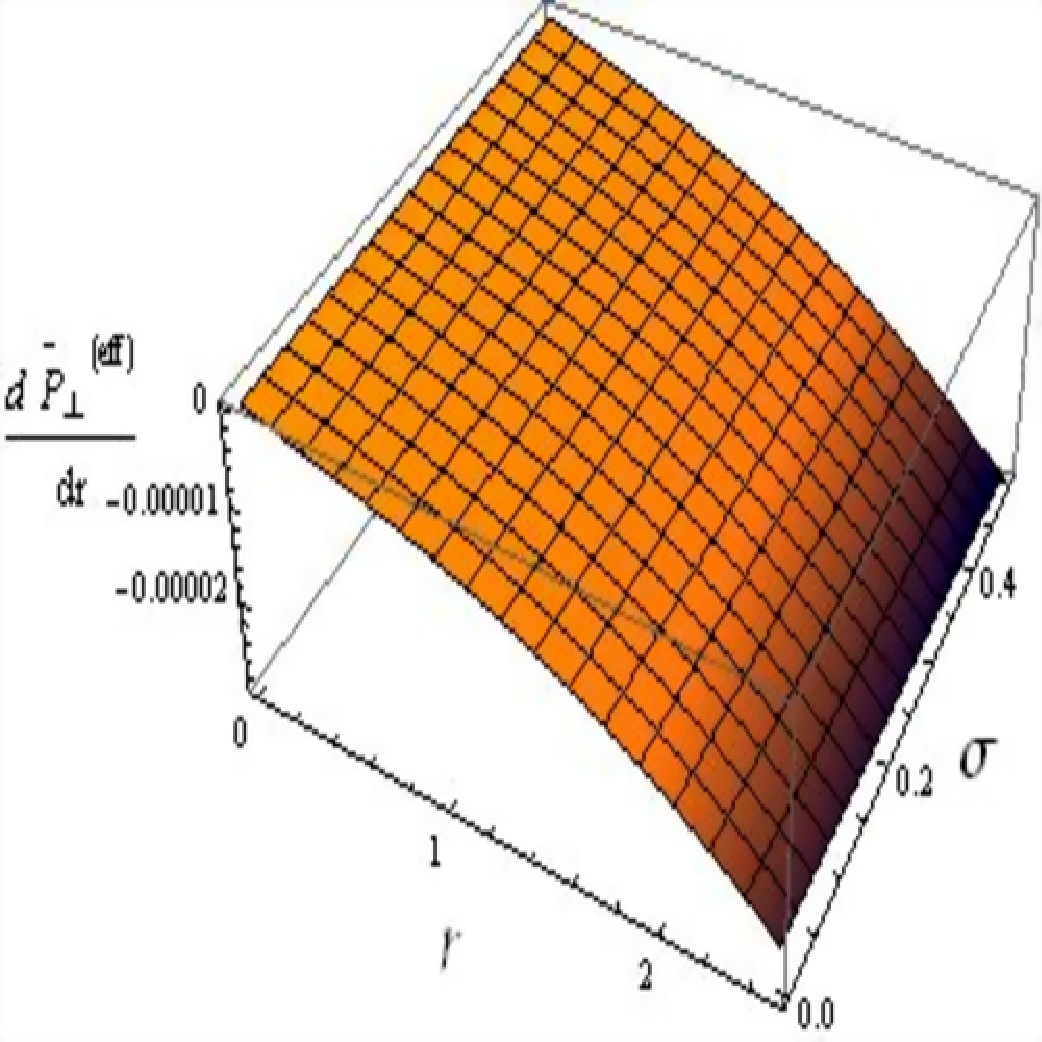,width=0.4\linewidth}\epsfig{file=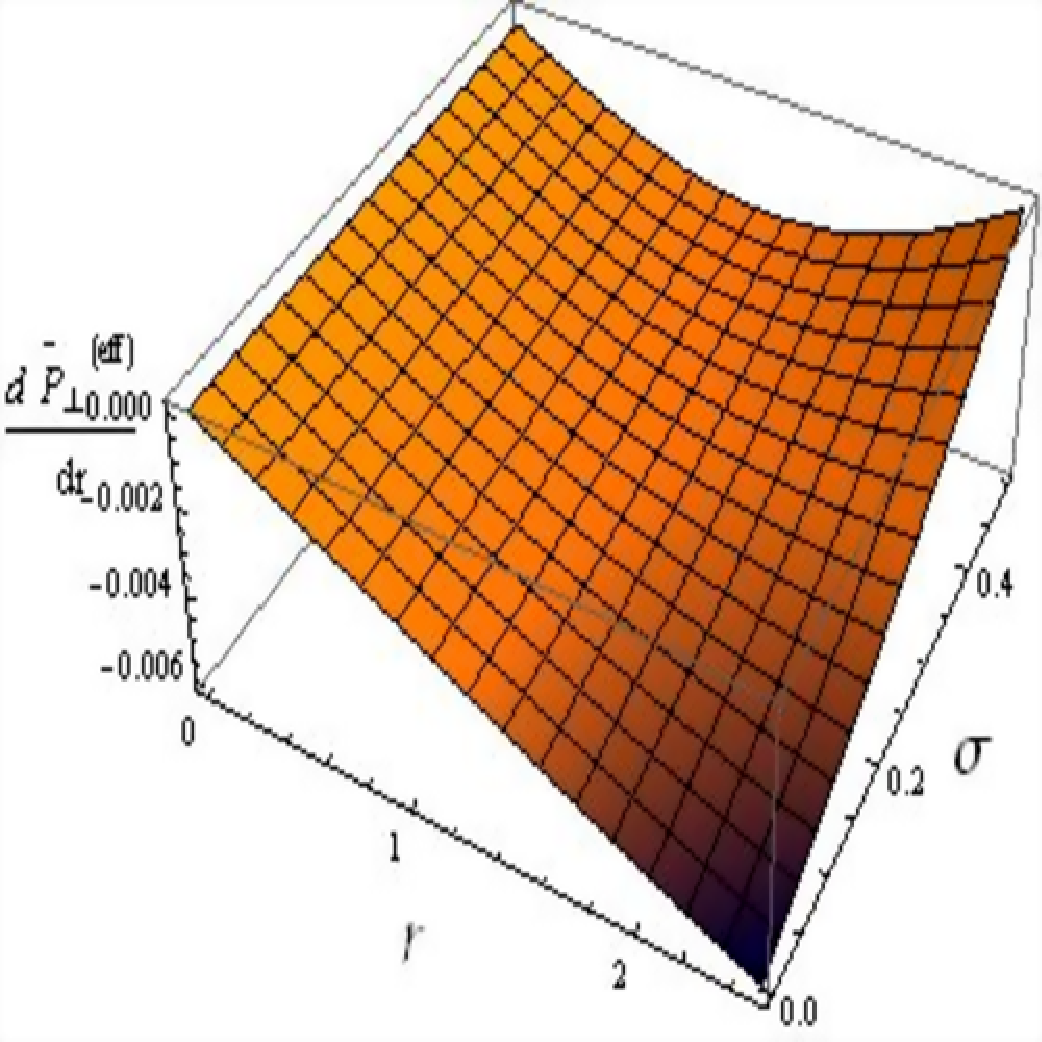,width=0.4\linewidth}
\caption{Plots of
$\frac{d\bar{\mu}^{(eff)}}{dr},~\frac{d\bar{P}_{r}^{(eff)}}{dr}$ and
$\frac{\bar{P}_{\bot}^{(eff)}}{dr}$ versus $r$ and $\sigma$
corresponding to solution-I (left) and solution-II (right).}
\end{figure}
\begin{figure}\center
\epsfig{file=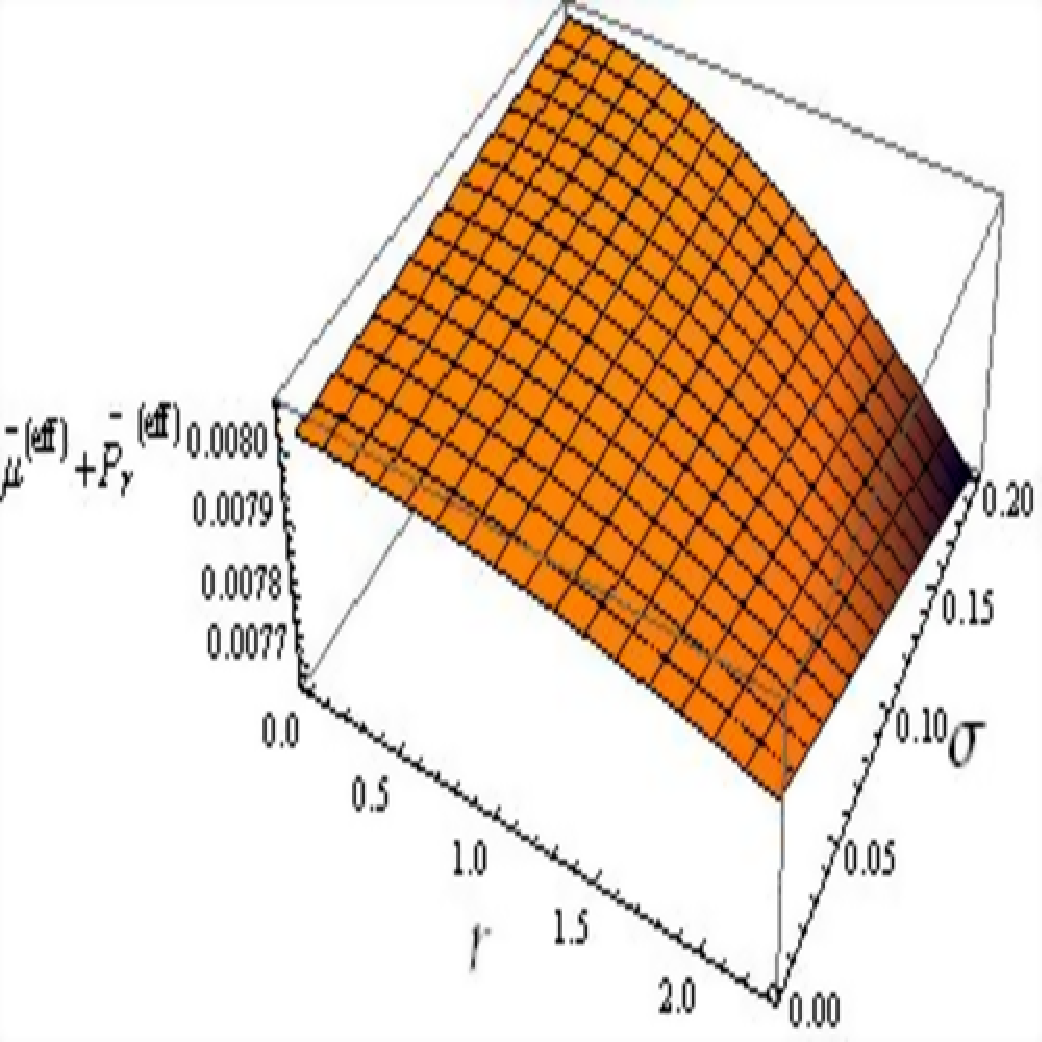,width=0.4\linewidth}\epsfig{file=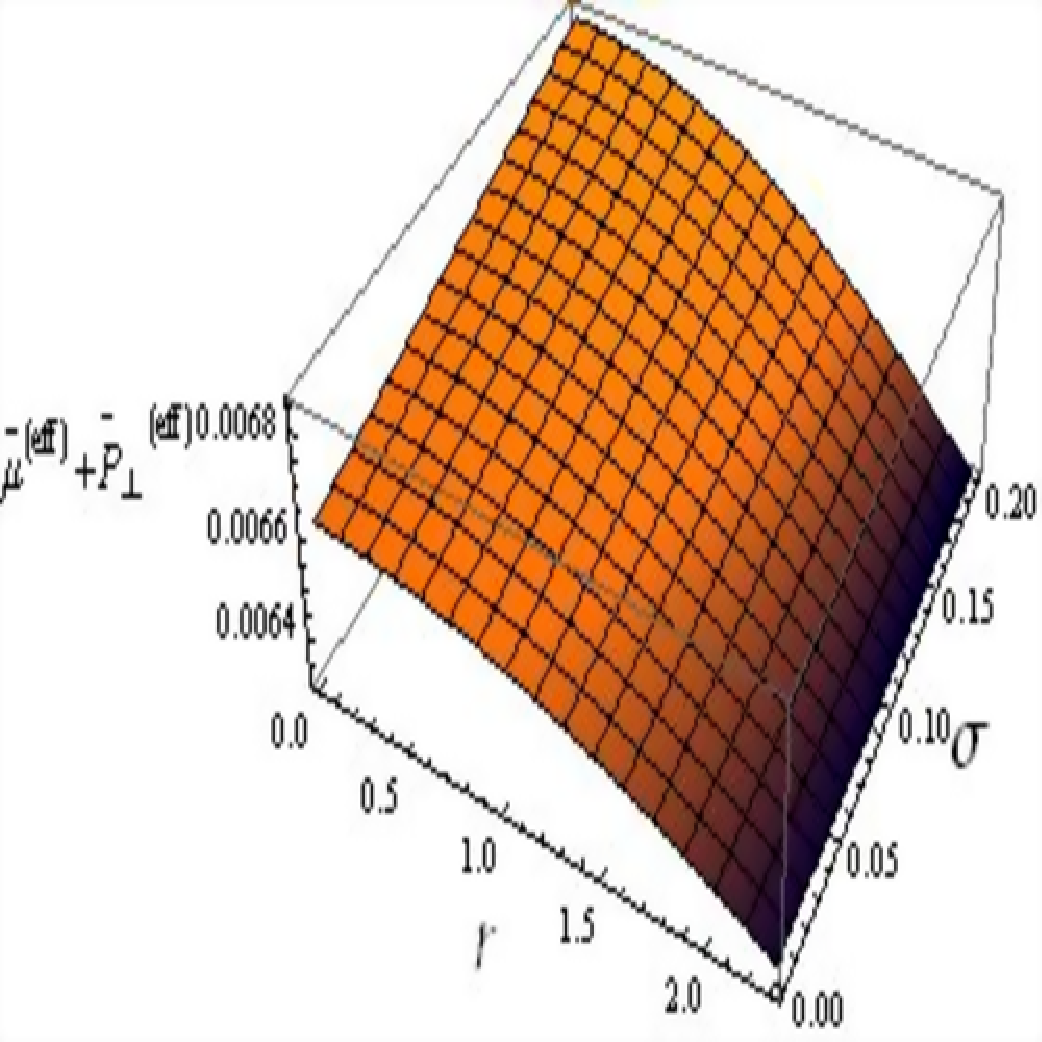,width=0.4\linewidth}
\epsfig{file=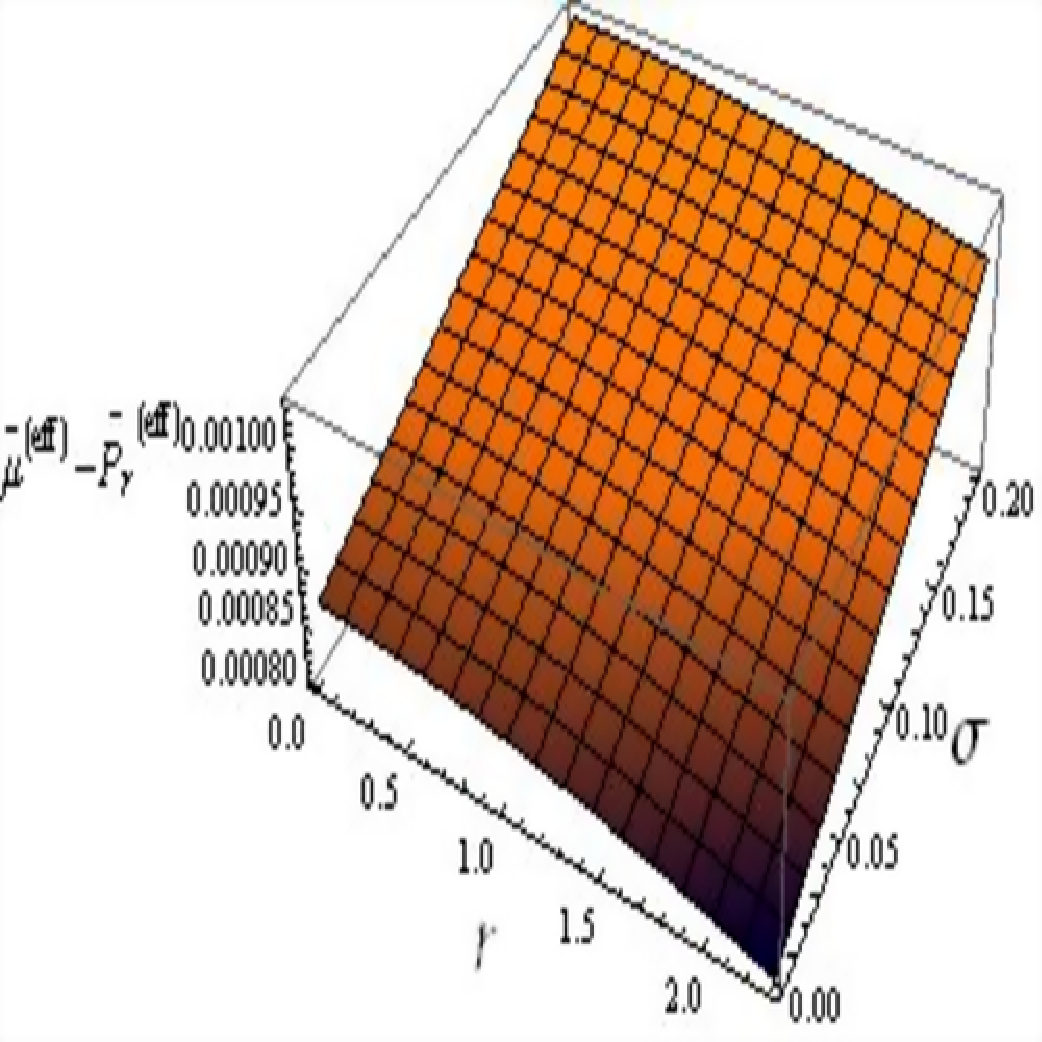,width=0.4\linewidth}\epsfig{file=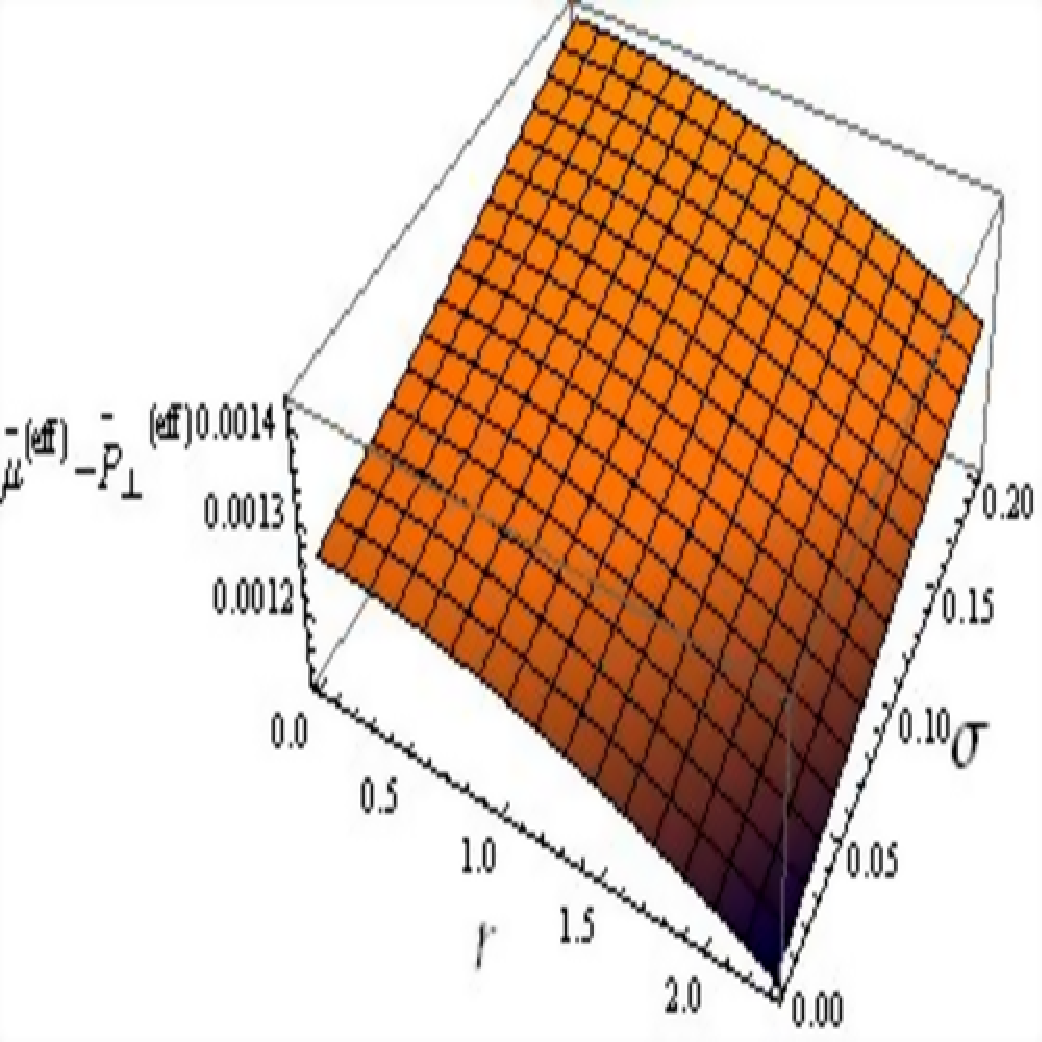,width=0.4\linewidth}
\epsfig{file=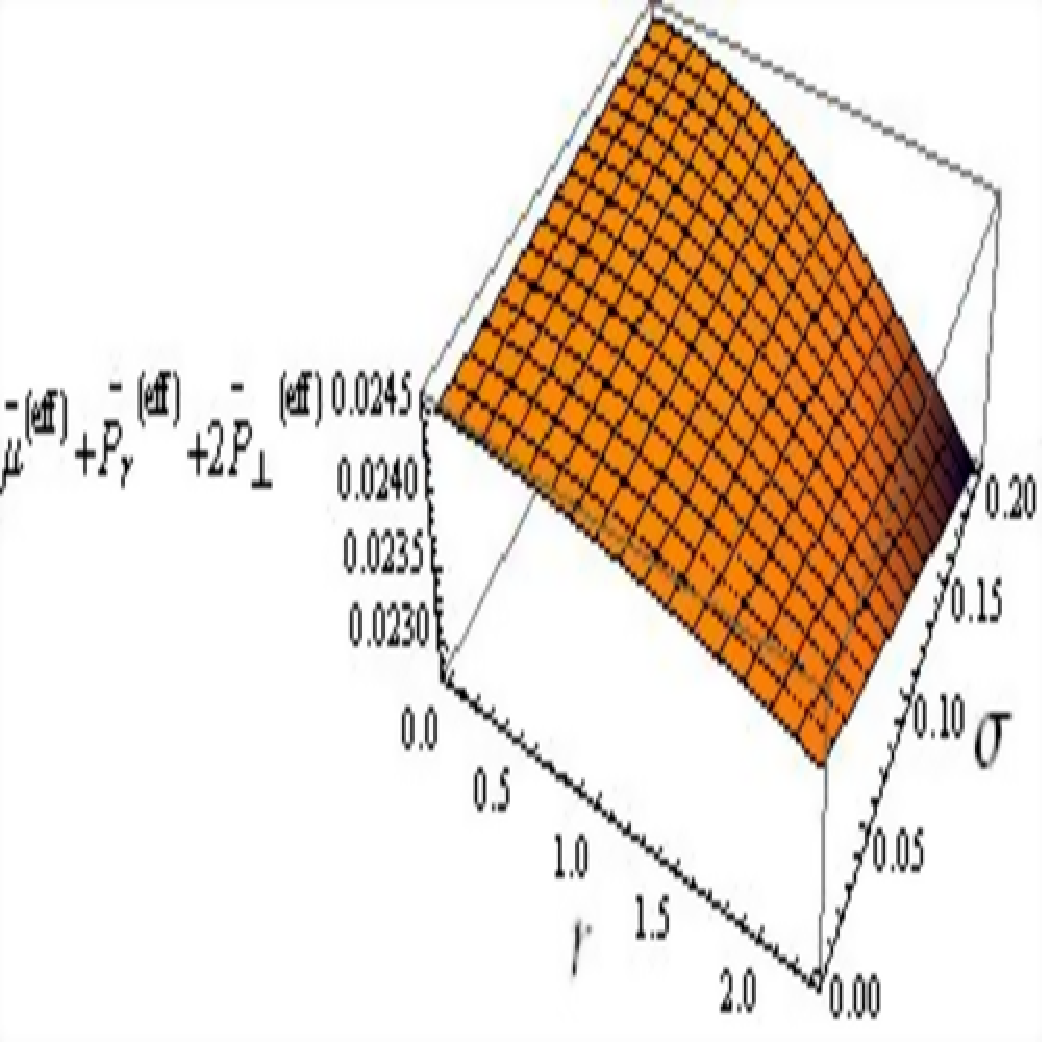,width=0.4\linewidth} \caption{Behavior of
energy bounds versus $r$ and $\sigma$ for solution-II.}
\end{figure}
\begin{figure}\center
\epsfig{file=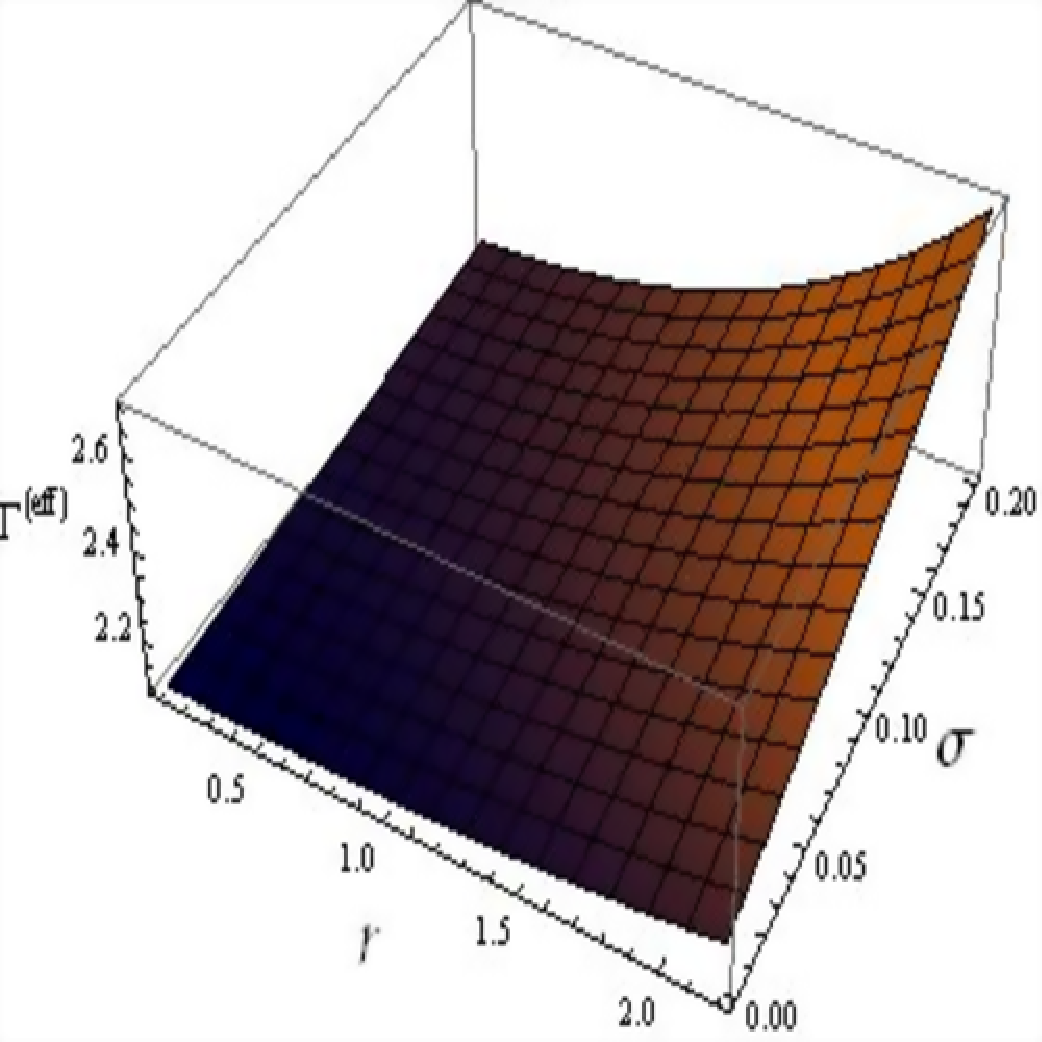,width=0.4\linewidth}\epsfig{file=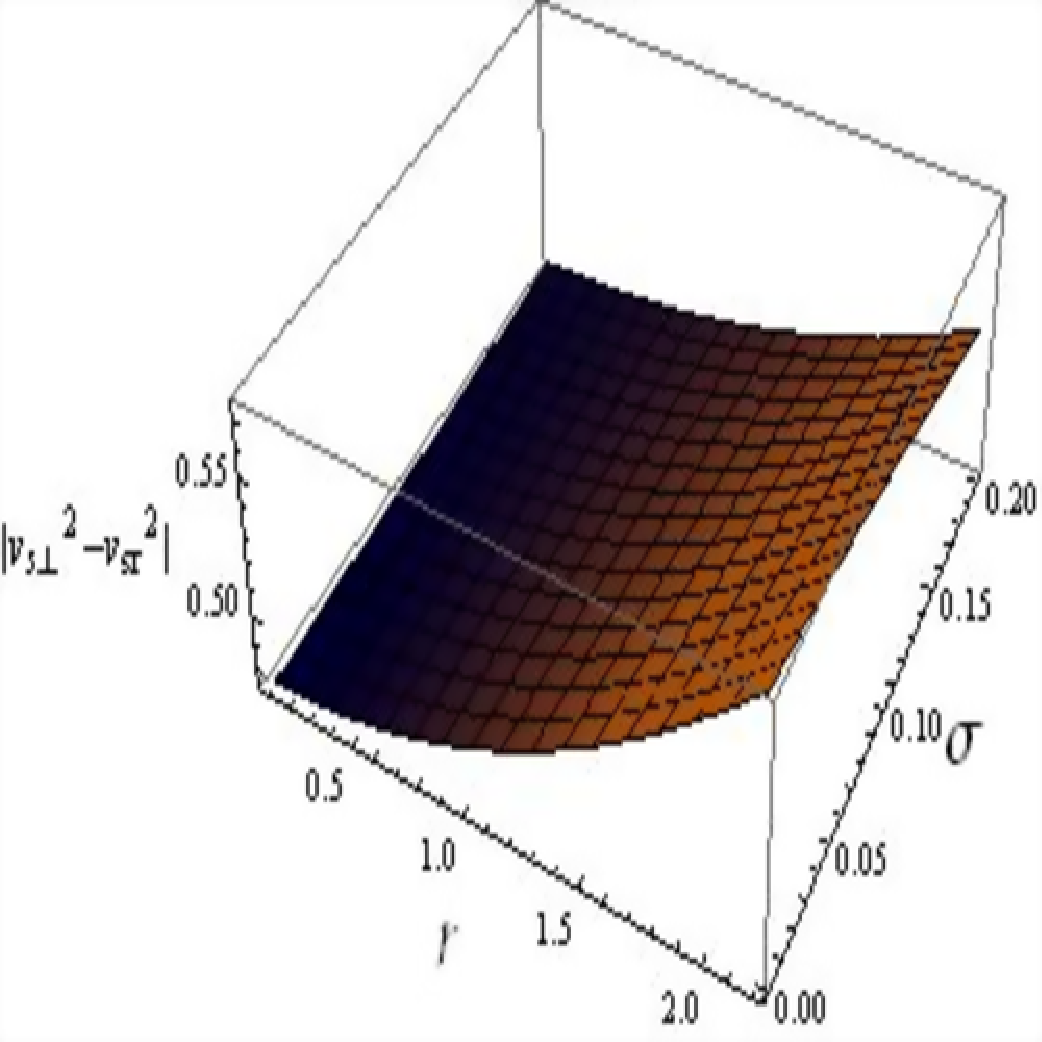,width=0.4\linewidth}
\caption{Plots of adiabatic index and $|v_{s\bot}^2-v_{sr}^2|$
versus $r$ and $\sigma$ for solution-II.}
\end{figure}
\begin{figure}\center
\epsfig{file=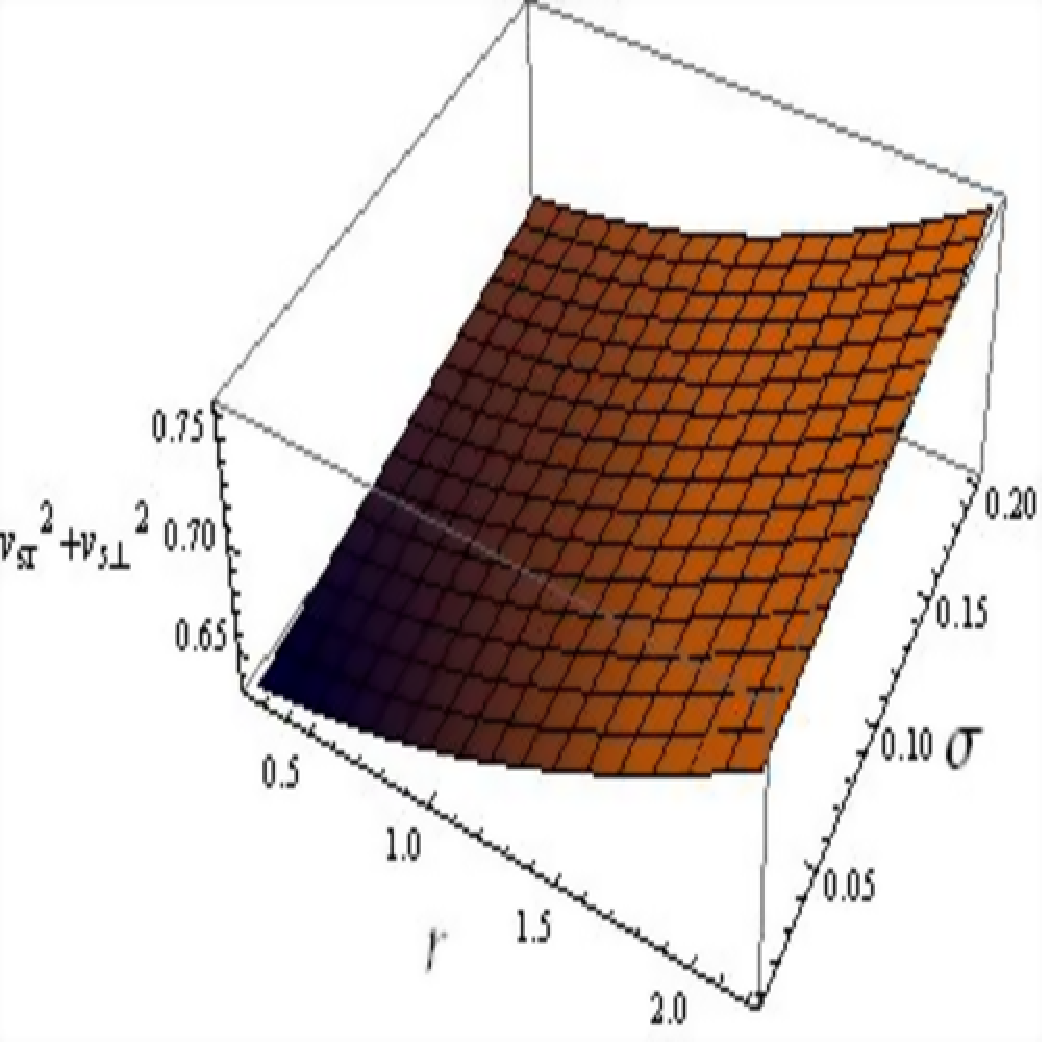,width=0.4\linewidth}\epsfig{file=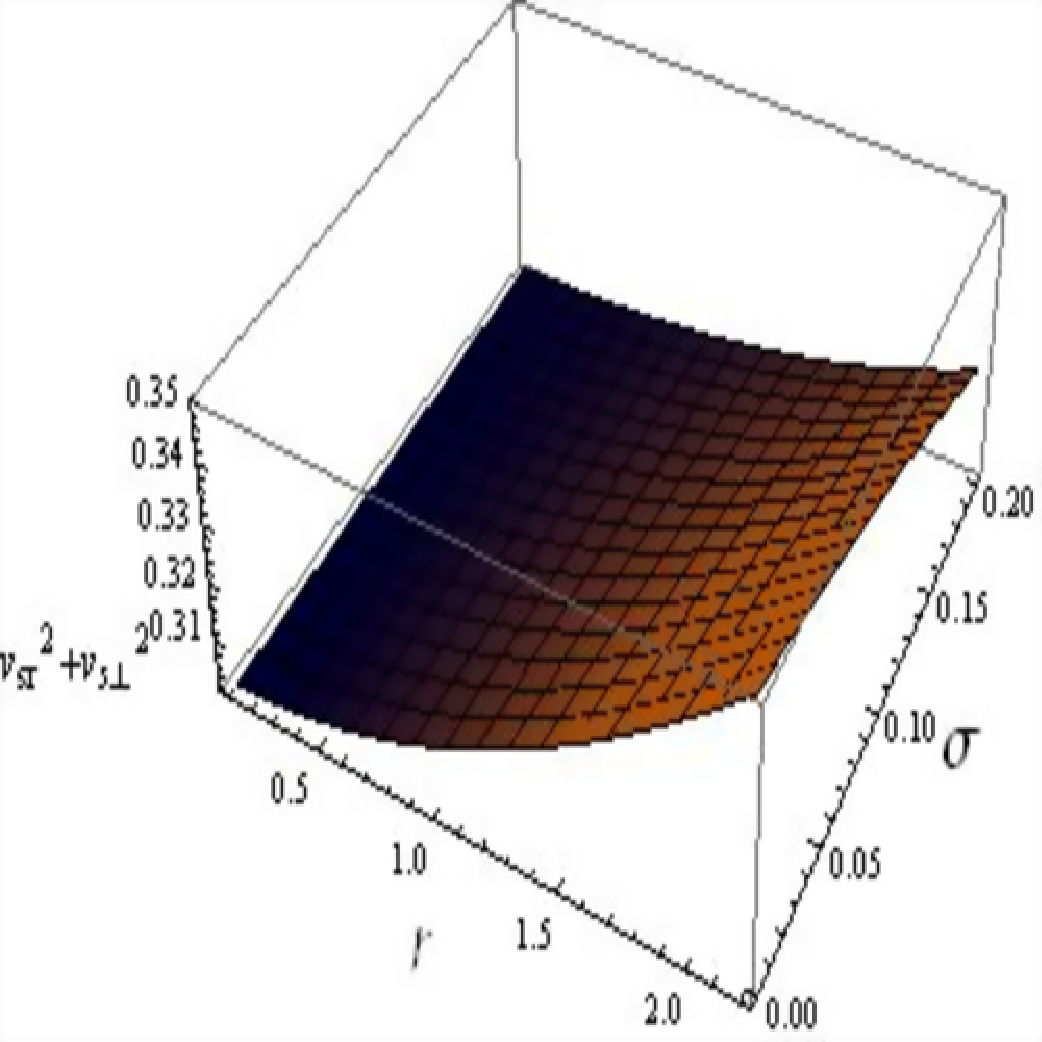,width=0.4\linewidth}
\caption{Plots of $v_{sr}^2+v_{s\bot}^2$ versus $r$ and $\sigma$ for
solution-I (left) and solution-II (right).}
\end{figure}

\section{Conclusions}

This paper is devoted to studying anisotropic spherical solutions of
self-gravitating object through gravitational decoupling technique
in $f(R,T,Q)$ theory. Here, we have used a linear model $R+\alpha Q$
of this curvature-matter coupled gravity. Two anisotropic solutions
have been obtained by adding an extra term $\Theta_{\rho\eta}$ in
the isotropic solution. We have taken the Krori-Barua ansatz and
determined unknown quantities by means of matching criteria. There
are four unknown quantities in the second sector
\eqref{g21}-\eqref{g23} which are reduced by implementing an extra
constraint on $\Theta_{\rho\eta}$.

We have utilized two constraints which equal the effective pressure
and energy density of the original isotropic distribution and
additional anisotropic source to develop solutions-I and II,
respectively. The physical behavior of state variables
$(\bar{\mu}^{(eff)},\bar{P}_{r}^{(eff)},\bar{P}_{\bot}^{(eff)})$,
anisotropy $(\bar{\Delta}^{(eff)})$ and energy conditions
\eqref{g50} are examined for $\alpha=-0.3$ to assess the acceptance
of these solutions. It is found that our both solutions fulfil the
needed limit for compactness and redshift. It is obtained that
stellar structure corresponding to the solution-I becomes more dense
for larger values of the decoupling parameter $\sigma$, whereas it
becomes less dense for the solution-II. The stability of the
resulting solutions has also been examined through cracking approach
and the adiabatic index. We have found that both solutions meet the
stability criteria and also physically viable as they fulfil the
energy bounds. It is worth mentioning here that our resulting
solutions are physically viable as well as stable for larger values
of $\sigma$ contrary to GR and $f(G)$ gravity \cite{34,35}. Thus,
this technique in $f(R,T,Q)$ gravity provides more suitable results.
Our results are consistent with $f(R)$ theory \cite{35a}. Finally,
we would like to mention here that all these findings reduce to GR
when $\alpha=0$ in the model \eqref{g61}.

\vspace{0.25cm}

\section*{Appendix A}

The matter components involving modified corrections appearing in
Eqs.\eqref{g8}-\eqref{g10} are
\begin{eqnarray}\nonumber
T_{0}^{0(D)}&=&\frac{1}{8\pi(f_{R}+\mu
f_{Q})}\left[\mu\left\{f_{Q}\left(\frac{\nu'\chi'}{4e^{\chi}}-\frac{\nu'}{re^{\chi}}+\frac{\nu'^2}{2e^{\chi}}
-\frac{\nu''}{2e^{\chi}}-\frac{1}{2}R\right)+f'_{Q}\left(\frac{\nu'}{2e^{\chi}}\right.\right.\right.\\\nonumber
&+&\left.\left.\frac{1}{re^{\chi}}-\frac{\chi'}{4e^{\chi}}\right)+\frac{f''_{Q}}{2e^{\chi}}-2f_{T}\right\}
+\mu'\left\{f_{Q}\left(\frac{\nu'}{2e^{\chi}}
-\frac{\chi'}{4e^{\chi}}+\frac{1}{re^{\chi}}\right)+\frac{f'_{Q}}{e^{\chi}}\right\}\\\nonumber
&+&\frac{f_{Q}\mu''}{2e^{\chi}}+P\left\{f_{Q}\left(\frac{3\chi'^2}{4e^{\chi}}-\frac{\chi''}{2e^{\chi}}-\frac{2}{r^2e^{\chi}}\right)
-f'_{Q}\left(\frac{5\chi'}{4e^{\chi}}-\frac{1}{re^{\chi}}\right)+\frac{f''_{Q}}{2e^{\chi}}\right\}\\\nonumber
&+&P'\left\{f_{Q}\left(\frac{1}{re^{\chi}}
-\frac{5\chi'}{4e^{\chi}}\right)+\frac{f'_{Q}}{e^{\chi}}\right\}+\frac{f_{Q}P''}{2e^{\chi}}+\frac{Rf_{R}}{2}
+f'_{R}\left(\frac{\chi'}{2e^{\chi}}-\frac{2}{re^{\chi}}\right)\\\nonumber
&-&\left.\frac{f''_{R}}{e^{\chi}}-\frac{f}{2}\right],\\\nonumber
T_{1}^{1(D)}&=&\frac{1}{8\pi(f_{R}+\mu
f_{Q})}\left[\mu\left(f_{T}-\frac{f_{Q}\nu'^2}{4e^{\chi}}
+\frac{f'_{Q}\nu'}{4e^{\chi}}\right)+\frac{f_{Q}\mu'\nu'}{4e^{\chi}}+P\left\{f_{T}
+f_{Q}\left(\frac{\nu''}{e^{\chi}}\right.\right.\right.\\\nonumber
&+&\left.\left.\frac{\nu'^2}{2e^{\chi}}-\frac{\chi'^2}{e^{\chi}}-\frac{3\chi'}{re^{\chi}}-\frac{3\nu'\chi'}{4e^{\chi}}
+\frac{2}{r^2e^{\chi}}+\frac{1}{2}R\right)
-f'_{Q}\left(\frac{\nu'}{4e^{\chi}}+\frac{2}{re^{\chi}}\right)\right\}\\\nonumber
&-&\left.P'f_{Q}\left(\frac{\nu'}{4e^{\chi}}+\frac{2}{re^{\chi}}\right)+\frac{f}{2}-\frac{Rf_{R}}{2}
-f'_{R}\left(\frac{\nu'}{2e^{\chi}}+\frac{2}{re^{\chi}}\right)\right],\\\nonumber
T_{2}^{2(D)}&=&\frac{1}{8\pi(f_{R}+\mu
f_{Q})}\left[\mu\left(f_{T}-\frac{f_{Q}\nu'^2}{4e^{\chi}}
+\frac{f'_{Q}\nu'}{4e^{\chi}}\right)+\frac{f_{Q}\mu'\nu'}{4e^{\chi}}+P\left\{f_{T}
+f_{Q}\left(\frac{\chi''}{2e^{\chi}}\right.\right.\right.\\\nonumber
&-&\left.\frac{3\chi'^2}{4e^{\chi}}+\frac{\nu'}{2re^{\chi}}-\frac{\chi'}{2re^{\chi}}-\frac{2}{r^2}+\frac{1}{r^2e^{\chi}}
+\frac{1}{2}R\right)+f'_{Q}\left(\frac{3\chi'}{2e^{\chi}}-\frac{3}{re^{\chi}}-\frac{\nu'}{4e^{\chi}}\right)\\\nonumber
&-&\left.\frac{f''_{Q}}{e^{\chi}}\right\}+P'\left\{f_{Q}\left(\frac{3\chi'}{2e^{\chi}}-\frac{3}{re^{\chi}}
-\frac{\nu'}{4e^{\chi}}\right)-\frac{2f'_{Q}}{e^{\chi}}\right\}-\frac{f_{Q}P''}{e^{\chi}}-\frac{Rf_{R}}{2}+\frac{f}{2}\\\nonumber
&+&\left.f'_{R}\left(\frac{\chi'}{2e^{\chi}}-\frac{\nu'}{2e^{\chi}}
-\frac{1}{re^{\chi}}\right)-\frac{f''_{R}}{e^{\chi}}\right].
\end{eqnarray}
The quantity $\Omega$ in Eq.\eqref{g12} is given as
\begin{align}
\nonumber \Omega &=
\frac{2}{\left(Rf_{Q}+2(8\pi+f_{T})\right)}\left[f'_{Q}e^{-\chi}P\left(\frac{1}{r^2}-\frac{e^\chi}{r^2}
+\frac{\nu'}{r}\right)+f_{Q}e^{-\chi}P\left(\frac{\nu''}{r}-\frac{\nu'}{r^2}-\frac{\chi'}{r^2}\right.\right.\\\nonumber
&-\left.\frac{\nu'\chi'}{r}-\frac{2}{r^3}+\frac{2e^\chi}{r^3}\right)+P'\left\{f_{Q}e^{-\chi}\left(\frac{\nu'\chi'}{8}
-\frac{\nu''}{8}-\frac{\nu'^2}{8}+\frac{\chi'}{2r}+\frac{\nu'}{2r}+\frac{1}{r^2}-\frac{e^{\chi}}{r^2}\right)\right.\\\nonumber
&+\left.\frac{3}{4}f_{T}\right\}+Pf'_{T}-\mu
f'_{T}-\mu'\left\{\frac{3f_{T}}{2}+\frac{f_{Q}e^{-\chi}}{8}\left(\nu'^2-\nu'\chi'+2\nu''+\frac{4\nu'}{r}\right)\right\}\\\nonumber
&+\left.\left(\frac{1}{r^2}-\frac{e^{-\chi}}{r^2}-\frac{\nu'e^{-\chi}}{r}\right)\left(\mu'f_{Q}+\mu
f'_{Q}\right)\right].
\end{align}
The adiabatic index corresponding to solutions-I and II are
\begin{align}
\nonumber
\Gamma^{(eff)}&=-\bigg[\big(2\mathcal{B}r^2\big(\pi\alpha\big(3\mathcal{A}r^2-7\big)-\sigma-1\big)
+4\pi\alpha\mathcal{A}r^2+(\sigma+1)e^{\mathcal{A}r^2}+2\pi\alpha\mathcal{B}^2r^4\\\nonumber
&-\sigma-1\big)\big(e^{\mathcal{A}r^2}\big(8\mathcal{B}^3r^6+4\mathcal{B}^2r^4(3-2\sigma)
+6\mathcal{B}r^2(\sigma+1)+\sigma+1\big)-\big(2\mathcal{B}r^2\\\nonumber
&+1\big)^3\big(-12\pi\alpha\mathcal{A}^3r^6+2\mathcal{A}^2r^4\big(3\pi\alpha\big(\mathcal{B}r^2+2\big)-\sigma-1\big)
+\mathcal{A}r^2\big(18\pi\alpha\mathcal{B}^2r^4\\\nonumber
&-28\pi\alpha\mathcal{B}r^2+\sigma+1\big)-18\pi\alpha\mathcal{B}^2r^4+\sigma+1\big)\big)\bigg]^{-1}
\bigg[2r^2\big(2\mathcal{B}r^2+1\big)\big(2\pi\alpha\mathcal{A}^2r^4\\\nonumber
&\times\big(3\mathcal{B}r^2+2\big)+\mathcal{A}r^2\big(2\pi\alpha\mathcal{B}^2r^4
-2\mathcal{B}r^2(10\pi\alpha+\sigma+1)-\sigma-1\big)+(\sigma+1)\\\nonumber
&\times
e^{\mathcal{A}r^2}-2\pi\alpha\mathcal{B}^2r^4-\sigma-1\big)\big(-6\pi\alpha\mathcal{A}^2\big(2\mathcal{B}r^3+r\big)^2
+\mathcal{B}\big(-18\pi\alpha-2\mathcal{B}r^2\\\nonumber
&\times\big(31\pi\alpha-\sigma
e^{\mathcal{A}r^2}+2\sigma+2\big)+3\sigma
e^{\mathcal{A}r^2}+40\pi\alpha\mathcal{B}^3r^6-4\mathcal{B}^2r^4(8\pi\alpha+\sigma+1)\\\nonumber
&-\sigma-1\big)+\mathcal{A}\big(2\mathcal{B}r^2+1\big)^2\big(2\pi\big(\alpha+3\alpha\mathcal{B}r^2\big)-\sigma-1\big)\big)\bigg],\\\nonumber
\Gamma^{(eff)}&=\bigg[\big(12\pi\alpha\mathcal{A}^3r^6+2\mathcal{A}^2r^4\big(-3\pi\alpha\big(\mathcal{B}r^2+2\big)+\sigma+1\big)
-\mathcal{A}r^2\big(18\pi\alpha\mathcal{B}^2r^4\\\nonumber
&-28\pi\alpha\mathcal{B}r^2+\sigma+1\big)+(\sigma+1)e^{\mathcal{A}r^2}+18\pi\alpha\mathcal{B}^2r^4-\sigma-1\big)
\big(2\mathcal{B}r^2\big(\pi\alpha\\\nonumber
&\times\big(3\mathcal{A}r^2-7\big)-\sigma-1\big)+e^{\mathcal{A}r^2}\big(2\mathcal{B}r^2\sigma+\sigma+1\big)
+4\pi\alpha\mathcal{A}r^2+2\pi\alpha\mathcal{B}^2r^4\\\nonumber
&-\sigma-1\big)\bigg]^{-1}\bigg[2
r^2\big(6\pi\alpha\mathcal{A}^2r^2+\mathcal{A}\big(-2\pi\big(\alpha+3\alpha\mathcal{B}r^2\big)+\sigma+1\big)+\mathcal{B}\big(-\sigma\\\nonumber
&\times
e^{\mathcal{A}r^2}-2\pi\alpha\big(5\mathcal{B}r^2-9\big)+\sigma+1\big)\big)\big(2\pi\alpha\mathcal{A}^2r^4\big(3\mathcal{B}r^2+2\big)
+\mathcal{A}r^2\big(2\pi\alpha\mathcal{B}^2r^4\\\nonumber
&-2\mathcal{B}r^2(10\pi\alpha+\sigma+1)-\sigma-1\big)+(\sigma+1)e^{\mathcal{A}r^2}-2\pi\alpha\mathcal{B}^2r^4-\sigma-1\big)\bigg].
\end{align}
The value of $|v_{st}^2-v_{sr}^2|$ corresponding to solutions-I and
II become
\begin{align}\nonumber
|v_{st}^2-v_{sr}^2|&=\bigg|\bigg[\big(2\mathcal{B}r^2+1\big)^3\big(-12\pi\alpha\mathcal{A}^3r^6+2\mathcal{A}^2r^4
\big(3\pi\alpha\big(\mathcal{B}r^2+2\big)-\sigma-1\big)\\\nonumber
&+\mathcal{A}r^2\big(18\pi\alpha\mathcal{B}^2r^4-28\pi\alpha\mathcal{B}r^2+\sigma+1\big)-18\pi\alpha\mathcal{B}^2r^4
+\sigma+1\big)-e^{\mathcal{A}r^2}\\\nonumber
&\times\big(8\mathcal{B}^3r^6+4\mathcal{B}^2r^4(3-2\sigma)+6\mathcal{B}r^2(\sigma+1)+\sigma+1\big)\bigg]^{-1}
\bigg[\big(2\mathcal{B}r^2+1\big)\\\nonumber
&\times\big(-\mathcal{B}^2r^4\big(-16\pi\alpha
+(16\pi\alpha-4\sigma+3)e^{\mathcal{A}r^2}+5\sigma\big)+\mathcal{A}r^2\big(4\pi\alpha+\mathcal{B}r^2\\\nonumber
&\times\big(16\pi\alpha+e^{\mathcal{A}r^2}-3\sigma\big)+4\mathcal{B}^4r^8\sigma+8\mathcal{B}^3r^6\sigma+\mathcal{B}^2r^4
(16\pi\alpha+\sigma)-\sigma\big)\\\nonumber
&-4\mathcal{B}r^2(4\pi\alpha-\sigma)\big(e^{\mathcal{A}r^2}-1\big)-r^4\sigma\big(\mathcal{B}r^2+1\big)
\big(2\mathcal{A}\mathcal{B}r^2+\mathcal{A}\big)^2
+(4\pi\alpha\\\nonumber
&-\sigma)\big(1-e^{\mathcal{A}r^2}\big)-4\mathcal{B}^4r^8\sigma-4\mathcal{B}^3r^6\sigma\big)\bigg]\bigg|,\\\nonumber
|v_{st}^2-v_{sr}^2|&=\bigg|\bigg[12\pi\alpha\mathcal{A}^3r^6+2\mathcal{A}^2r^4\big(-3\pi\alpha\big(\mathcal{B}r^2+2\big)+\sigma+1\big)
-\mathcal{A}r^2\big(18\pi\alpha\mathcal{B}^2r^4\\\nonumber
&-28\pi\alpha\mathcal{B}r^2+\sigma+1\big)+(\sigma+1)e^{\mathcal{A}r^2}+18\pi\alpha\mathcal{B}^2r^4-\sigma-1\bigg]^{-1}
\bigg[\sigma\big(\mathcal{A}^2\big(\mathcal{B}r^6\\\nonumber
&+r^4\big)+e^{\mathcal{A}r^2}\big(\mathcal{A}\mathcal{B}r^4-\mathcal{B}^2r^4-1\big)+\mathcal{A}\big(-\mathcal{B}^2r^6-\mathcal{B}r^4+r^2\big)
+\mathcal{B}^2r^4+1\big)\\\nonumber
&+4\pi\alpha\big(-\mathcal{A}r^2+e^{\mathcal{A}r^2}-1\big)\bigg]\bigg|.
\end{align}

\end{document}